\documentclass[12pt]{article}
\pdfoutput=1
\usepackage{latexsym, graphicx,cite}
\usepackage{amsmath}
\usepackage{amssymb}
\usepackage{amsthm}
\usepackage{braket}
\numberwithin{equation}{section}

\allowdisplaybreaks

\usepackage[top = 1. in,bottom = 1. in, left = 1 in, right=1 in]{geometry}

\newcommand{\arXiv}[1]{\href{http://www.arXiv.org/abs/#1}{arXiv:#1}}
\usepackage[colorlinks=true, linkcolor=blue, bookmarks=true]{hyperref}

\usepackage{cleveref}
\crefname{equation}{}{}

\makeatletter
\renewcommand\section{\@startsection {section}{1}{\z@}%
   {-3.5ex \@plus -1ex \@minus -.2ex}
   {2.3ex \@plus.2ex}%
   {\normalfont\large\bfseries}}
\renewcommand\subsection{\@startsection{subsection}{2}{\z@}%
   {-3.25ex\@plus -1ex \@minus -.2ex}%
   {1.5ex \@plus .2ex}%
   {\normalfont\bfseries}}
\crefrangelabelformat{equation}{%
  \protected@edef\eq@ref@a{#1}%
  \protected@edef\eq@ref@b{#2}%
  \expandafter\expandafter\expandafter
  \eq@ref@check\expandafter\eq@ref@a\expandafter.\expandafter\@nil
                           \eq@ref@b.\@nil
  (%
    #3\eq@ref@a#4%
    --%
    #5\eq@ref@b#6%
  )%
}
\def\eq@ref@check#1.#2\@nil#3.#4\@nil{%
  \def\eq@tmp{#2}%
  \ifx\eq@tmp\@empty
  \else
    \def\eq@tmp{#4}%
    \ifx\eq@tmp\@empty
    \else
      \def\eq@tmp@a{#1}%
      \def\eq@tmp@b{#3}%
      \ifx\eq@tmp@a\eq@tmp@b
        \expandafter\def\expandafter\eq@ref@b\expandafter{%
          \eq@strip@dot#4\@nil
        }%
      \fi 
    \fi
  \fi
}
\def\eq@strip@dot#1.\@nil{#1}
\makeatother

\newcommand{\beq}{\begin{equation}}
\newcommand{\eeq}{\end{equation}}
\newcommand{\ber}{\begin{array}}
\newcommand{\eer}{\end{array}}
\newcommand{\del}{\partial}

\newcommand{\dsty}{\displaystyle}

\newcommand{\te}{\theta}
\newcommand{\ph}{\varphi}
\newcommand{\de}{\delta}
\newcommand{\De}{\Delta}

\newcommand{\eps}{\varepsilon}
\newcommand{\al}{\alpha}
\newcommand{\alb}{\bar{\alpha}}

\newcommand{\ab}{\bar\alpha}

\newcommand{\om}{\omega}

\newcommand{\indices}{nlk}
\newcommand{\ei}{e_{nlk}}
\newcommand{\oi}{\omega_{nlk}}
\newcommand{\G}[1]{\Gamma{\left(#1\right)}}
\newcommand{\B}[2]{B\left(#1,#2\right)}
\newcommand{\be}{\beta}
\newcommand{\beb}{\bar{\beta}}

\newcommand{\bei}{\beta_{\indices}}
\newcommand{\bb}{\bar{b}}
\newcommand{\ena}{\end{eqnarray}}
\newcommand{\beqa}{\begin{eqnarray}}
\newcommand{\eeqa}{\end{eqnarray}}
\newcommand{\bea}{\begin{eqnarray}}
\newcommand{\eea}{\end{eqnarray}}

\newcommand{\yd}{\dot{y}}
\newcommand{\ali}{\alpha_{\indices}}

\renewcommand{\Re}{\operatorname{Re}}
\renewcommand{\Im}{\operatorname{Im}}

\begin{document}
\begin{titlepage}
\begin{flushright}
\phantom{arXiv:yymm.nnnn}
\end{flushright}
\vspace{1cm}
\begin{center}
{\LARGE\bf Time-periodicities in holographic CFTs}\\
\vskip 15mm
{\large Ben Craps,$^a$ Marine De Clerck$^{\,a}$ and Oleg Evnin$^{b,a}$}
\vskip 7mm
{\em $^a$ Theoretische Natuurkunde, Vrije Universiteit Brussel (VUB) and\\
The International Solvay Institutes, Brussels, Belgium}
\vskip 3mm
{\em $^b$ Department of Physics, Faculty of Science, Chulalongkorn University,
Bangkok, Thailand}
\vskip 7mm
{\small\noindent {\tt Ben.Craps@vub.be, Marine.Alexandra.De.Clerck@vub.be, oleg.evnin@gmail.com}}
\vskip 10mm
\end{center}
\vspace{1cm}
\begin{center}
{\bf ABSTRACT}\vspace{3mm}
\end{center}
Dynamics in AdS spacetimes is characterized by various time-periodicities. The most obvious of these is the time-periodic evolution of linearized fields, whose normal frequencies form integer-spaced ladders as a direct consequence of the structure of representations of the conformal group. There are also explicitly known time-periodic phenomena on much longer time scales inversely proportional to the coupling in the weakly nonlinear regime. We ask what would correspond to these long time periodicities in a holographic CFT, provided that such a CFT reproducing the AdS bulk dynamics in the large central charge limit has been found. The answer is a very large family of multiparticle operators whose conformal dimensions form simple ladders with spacing inversely proportional to the central charge. We give an explicit demonstration of these ideas in the context of a toy model holography involving a $\phi^4$ probe scalar field in AdS, but we expect the applicability of the underlying structure to be much more general.
\vfill

\end{titlepage}
{ \hypersetup{hidelinks} \tableofcontents }\vspace{5mm}
\newpage

\section{Introduction}

The AdS/CFT correspondence provides an attractive framework to connect the algebraic structures of conformal field theories (CFTs) with the dynamics of fields and strings in Anti-de Sitter (AdS) spacetimes. The precise connection between specific features of interest on the two sides of the correspondence may, however, be far from obvious. It is our goal here to explore a range of such connections, in particular, in relation to dynamical time-periodicity.

While the subject of CFTs has a long and accomplished history, studies of nonlinear dynamics in AdS spacetimes are relatively recent, largely triggered by the observations of \cite{BR} that suggested formation of black holes starting from arbitrarily small perturbations of some shapes in the initial state of the AdS evolution. A crucial ingredient underlying the sophistication of nonlinear dynamics in AdS for arbitrarily small perturbation amplitudes is the perfectly resonant spectrum of linearized normal mode frequencies. Thus for linear fields of any spin and mass in AdS backgrounds, the differences of frequencies of any two normal modes are integer in appropriate units (fixed in terms of the AdS curvature radius). Hence, pronounced time-periodic features appear already at the linearized levels (for example, all solutions of equations of motion for linear massless fields in AdS backgrounds are exactly periodic functions of time). Resonances between the linear normal modes are what enables nonlinearities to exert significant influences on perturbations of arbitrarily small amplitudes, provided that one waits long enough. A selection of further studies of this type of nonlinear dynamics can be found in \cite{AdSLMS,FPU,CEV,CEV2,BKS,revival,DKFK,DFLY,BMR,GMLL,BR2,CDSW,returns,AOQ,squashed}; for a short review,\footnote{While AdS instability remains an evidence-based conjecture for classical field systems in AdS of more direct relevance for the AdS/CFT program and hence for the present work, analogous effects have been rigorously established in mathematical studies of the Einstein-Vlasov system describing null dust in AdS \cite{Moschidis1,Moschidis2}.} see \cite{rev2}.

There are many other ways in which time periodicities, exact and approximate, manifest themselves in AdS dynamics. First of all, as one moves away
from the linearized regime, the associated time-periodic bounces with waves moving between the AdS interior and the boundary are retained in a distorted form.
Such bounces have been studied in numerical simulations at finite perturbation amplitudes in \cite{AdSLMS,revival}. In this case, the returns to the initial configuration are, of course, inexact, but they persist to appreciable perturbation amplitudes. Collapse into a black hole may happen after a number of such bounces, terminating the time-periodic evolution, but each individual bounce before that is approximately time-periodic.

Coming back to small perturbations, the effect of nonlinearities is to induce slow modulations of the time-periodic linearized dynamics. These slow modulations may accumulate to large effects due to the resonant nature of the linearized spectrum. For perturbations of amplitude $\eps$ and quartic interactions, the relevant
time scale for these modulations is $1/\eps^2$, often referred to as the `slow time'. These slow modulations may themselves be approximately or exactly time-periodic, with correspondingly large periods of order $1/\eps^2$. Thus, while the exact returns to the initial configuration seen in the linearized theory are upset by the effects of nonlinearities, the resulting nonlinear dynamics may still reconstruct the initial state very accurately, or even exactly, after a much longer waiting period. For gravitational perturbations of AdS$_4$, the initial state is reconstructed with extremely high precision (but imperfectly) after many bounces of the waves between the interior and the boundary, as has been observed in \cite{FPU} and investigated in detail in \cite{returns}, in analogy to the Fermi-Pasta-Ulam recurrence phenomena of nonlinear oscillator chains \cite{FPUrev}.

The situation is particularly striking when the modulations of linearized dynamics induced by nonlinearities reconstruct the initial state with perfect precision after a long waiting period. Such situations have been observed in the literature for non-gravitating self-interacting fields in non-dynamical AdS backgrounds \cite{CF,BEL}. As we shall explain below, one should expect similar phenomena for gravitating fields, but only outside spherical symmetry, where the analysis is computationally very challenging and has not been performed up to this date. For this reason, our main focus in this article will be on the dynamics of non-gravitating `probe' fields, keeping in mind that a generalization of our results to gravitating fields is likely, but would require substantial computational breakthroughs.

It is useful to keep in mind the mathematical relation between the nonlinear dynamics in AdS and simpler nonlinear Schr\"odinger equations
in harmonic potentials, which are used to describe trapped Bose-Einstein condensates in contemporary terrestrial experiments \cite{BDZ}. This relation was originally pointed out\footnote{Analogies between turbulent dynamics in AdS and nonlinear Schr\"odinger equations have been put forth even earlier \cite{DHMS} with reference to the much-studied nonlinear Schr\"odinger equation on a torus \cite{nls}. However, it is specifically after introducing a harmonic potential that the nonlinear Schr\"odinger equation develops a precise mathematical relation to the AdS problems.} in \cite{BMP} and explained by taking a nonrelativistic limit of AdS field equations in \cite{BEL}. The linearized dynamics in this case is described by the well-known equispaced perfectly resonant spectrum of the harmonic oscillator, while the weakly nonlinear dynamics displays direct parallels to the more complicated AdS case. This dynamics has been studied in \cite{BBCE,BBCE2} and shows some perfect periodic returns to the initial state. One may also take a nonrelativistic limit of gravitating systems in AdS, obtaining a Hartree equation in a harmonic potential, which also displays some perfect periodic returns \cite{BEF}. The underlying mathematical structure responsible for these weakly nonlinear periodic behaviors, common for AdS fields and their nonrelativistic harmonically trapped limits, has been made manifest in \cite{solvable,breathing}.

What is the CFT counterpart of all these time-periodic features in AdS dictated by the AdS/CFT correspondence?
When gravitating perturbations of AdS collapse to form black holes, this is seen as the AdS counterpart of thermalization in the dual CFT. When black holes fail to form, or are only formed after a long train of revivals
of the initial states, as happens for the approximately time-periodic scenarios, this is seen as an obstruction to effective thermalization in the dual CFT. Such a perspective
is developed in \cite{AdSLMS,FPU, revival} among other publications. But can one go beyond such `macroscopic' statements in the spirit of non-equilibrium statistical physics, and recover specific features in the evolution of `microscopic' CFT states that provide a counterpart to the classical AdS behaviors?

Here we come to a subtle point. The time periodicities we have described are for classical nonlinear fields in AdS, and hence finding a `microscopic' counterpart to this evolution on the CFT side requires identifying CFT states dual to semiclassical bulk geometries. This question does not appear
to be fully systematically settled, though connections between coherent states in CFTs and semiclassical configurations in the dual bulk have been put forward in \cite{coherent}.

Rather than systematically identifying the CFT counterparts of the classical configurations responsible for the time-periodic behaviors in AdS, however,
we prefer to turn things around and quantize our dynamics in the bulk. It will turn out that the weakly nonlinear dynamics of perturbations with amplitudes $\eps$ on times scales of order $1/\eps^2$ (or at coupling $\lambda$ on time scales $1/\lambda$) precisely corresponds to the lowest order nonlinear corrections to the energy eigenstates of quantum fields in AdS. As one stays at first order in the coupling parameter, no ultraviolet problems of quantum field theories arise, even if one chooses to include gravitational interactions. (Subtleties exist in defining interacting quantum field theories in AdS spacetimes beyond the leading order in the coupling parameter, see for instance  the recent article \cite{AdSQFT}, but they will not affect our considerations.) While straightforward to formulate and well-posed, the problem of finding these energy shifts is in general rather demanding
because of the huge (arbitrarily large) degeneracies of the quantum levels of free fields in AdS. By the standard lore of quantum-mechanical
perturbation theory for a degenerate spectrum, one has to diagonalize very large matrices made of the matrix elements of the interaction Hamiltonian
within each highly degenerate unperturbed level. Finding patterns in this diagonalization problem and connecting them to the classical time-periodic behaviors will form the technical core of our study.

Once our time-periodic behaviors in the bulk have been recast as properties of energy eigenstates of quantum fields in AdS at linear order in the coupling parameter, connecting them to the CFT side of the holographic duality becomes straightforward. Indeed, energy eigenstates in the AdS bulk are in one-to-one correspondence with the CFT operators of definite conformal dimension, as dictated by the conformal symmetry.
More specifically, discussions of small corrections to the conformal dimensions of the CFT operators are common in the context of `large $c$ holography'
 \cite{largec1,largec2}. In this picture, one assumes to have a family of CFTs with the central charge $c$ growing without a bound. In the limit $c\to\infty$, 
the conformal dimensions of the primary operators tend to infinity, except for a finite set of operators that acquire the properties of `generalized free fields' \cite{largec2}. Such fields, while their dimensions do not necessarily take the values corresponding to free fields, behave as free fields in the sense
that their correlators factorize. As a result, the conformal dimensions of such fields simply add up under taking products. If one considers the entire set of such product operators (often called the `multiparticle' operators), their conformal dimensions form a tower precisely corresponding to the energy
levels of a free quantum field in AdS. (A concrete realization of this picture in two-dimensional CFTs is known due to \cite{hspin,hspin2}.)
As one moves away from the strict $c=\infty$ limit, the dimensions receive corrections of order $1/c$ (and in particular, the highly degenerate
levels of conformal dimensions split due to these corrections). In the AdS bulk, this corresponds to the shifts of energy levels of free fields due to weak interaction, precisely the framework at which we arrived in the previous paragraph.

Analysis of the conformal dimensions of CFT operators is a common subject in the AdS/CFT literature. In application to self-interacting scalars, 
which will be our main focus in the technical part of the paper, these have been studied, for instance, in \cite{largec1, BSS,anomalous}.
The dimensions are typically analyzed in the language of correlation functions, which is different (albeit equivalent) to our presentation
(see, however, \cite{FSh} for an approach based on the Hamiltonian perturbation theory, which is much closer to our study).
A more important difference between our analysis and the bulk of the literature is that the latter typically focuses on `small' operators (made
of products of a few single-particle operators, and corresponding to states with just a few particles in the AdS bulk), which are often analyzed at higher orders
in the coupling parameter. By contrast, we shall always remain at the lowest nontrivial order in the coupling parameter, but are interested
in analyzing the corrections to conformal dimensions to arbitrarily `large' operators (corresponding to states with arbitrarily many particles in the bulk). This amounts to studying the diagonalization properties of a family of arbitrarily large, highly structured matrices.

In view of the above premises, what we have to do technically is to consider quantum fields in AdS and analyze the corrections to the energies of Hamiltonian eigenstates at leading order in the interaction strength, in particular, for high energy levels which are hugely degenerate in the noninteracting theory. In practice, we shall focus on a non-gravitating $\phi^4$ scalar within the maximally rotating sector (states that have the maximal possible amount of angular momentum for a given energy). We shall explain, following \cite{breathing}, that similar patterns should be expected in maximally rotating
sectors of gravitating systems, but recovering them explicitly would require substantial technical work beyond the scope of our treatment. As outlined already in \cite{madagascar}, the problem of finding these energy shifts can be reduced to diagonalizing a specific {\it quantum resonant system} \cite{quantres}, whose Hamiltonian is a quartic combination of creation-annihilation operators. Such quantum resonant systems are, on the one hand,
related to bosonic embedded Gaussian ensembles of random matrix theory \cite{Kota}, albeit without randomness in the couplings, and on the other hand, can be seen as a bosonic analog of the SYK model \cite{Kitaev, SYK1,SYK2} that has attracted much attention in the context of gravitational holography.

While the quantum resonant systems corresponding to our cases of interest cannot be fully solved analytically, our goal is to present their partial analytic solution:
a subset of energy levels and their explicit wavefunctions. This solution builds on the previous work \cite{split,timeLLL} for the simpler nonrelativistic analogs of the AdS systems. The explicit energy levels given by our solutions form simple ladders and provide clear quantum counterparts of the time-periodic
behaviors of the classical theory. We note that the structure that we find is by no means guaranteed to exist from the onset. Indeed, classical behaviors emerge from the high-energy asymptotics of the spectrum of the corresponding quantum system, and classical time-periodicities, even if exact, do not, in principle, have to originate from any exact properties of quantum eigenstates at finite energies. Despite this general observation, what we find in fact is a large family of explicit quantum energy eigenstates with a simple formula for the energy that not only explains the classical time periodicities, but also demonstrates a considerably larger amount of explicit tractable analytic structure in the quantum theory than what had been previously seen in its classical counterpart!

While our analysis was initially motivated by observing time-periodic behaviors in classical AdS dynamics, the value of the underlying explicit structure
in the corresponding energy levels that we find goes beyond this initial motivation and stands in its own right. In particular, it gives a prediction regarding properties of $1/c$ corrections to the conformal dimensions of multiparticle operators in a dual holographic CFT. It is nonetheless worthwhile, given our initial motivation, to make the connection between the family of quantum `ladder' states we find and the corresponding time-periodic classical solutions
as explicit as possible. To this end (again, building on the analysis of \cite{timeLLL} for the considerably simpler nonrelativistic case), we develop
a construction of coherent-like states made entirely out of our ladder states (and not involving any other energy eigenstates of the Hamiltonian)
that can approximate the time-periodic classical dynamics with arbitrary precision. This completes the circle and answers the initial question that had triggered this study, in addition to the explicit identification of the quantum level structure and its holographic CFT counterpart. The idea that coherent combinations of CFT eigenstates should correspond to semiclassical dynamics in the bulk has appeared in the literature before, see for example \cite{coherent} where ordinary harmonic oscillator states are employed in this manner. The states we construct are similar in spirit, but much more powerful, since, being made of our explicit ladder eigenstates of the Hamiltonian, they are adapted to the specific dynamics due to nonlinear interactions, and their evolution under the full interacting Hamiltonian at leading order in the interaction strength is completely straightforward.

Our exposition is organized as follows: In Section~\ref{sec:class}, we review classical perturbative treatments of systems with resonant spectra of normal mode frequencies, and applications of these methods to classical fields in AdS. Section~\ref{sec:quantum} discusses the fate of these structures under quantization. In Section~\ref{sec:large}, we review the basics of large $c$ holography, generalized free fields, towers of conformal dimensions of composite operators at $c=\infty$, and $1/c$ corrections to these towers, which are in direct correspondence with the quantum perturbation theory of section~\ref{sec:quantum}. In section~\ref{sec:rational}, we specialize to the maximally rotating sector of a $\phi^4$ scalar field in AdS and display an explicit `ladder' pattern in the $1/c$ corrections to the conformal dimensions in this sector. In section~\ref{sec:coherent}, we explain how to construct coherent-like combinations of the ladder states of section~\ref{sec:rational} that are in immediate correspondence with the classical weakly nonlinear solutions displaying time periodicity covered in section~\ref{sec:class}. We conclude in section~\ref{sec:discussion} with a discussion and point out further settings where similar structures are likely to emerge, beyond probe fields in AdS.


\section{Classical dynamics in AdS and its time periodicities}\label{sec:class}

We shall start by reviewing the background material necessary for our subsequent technical investigations. First of all, we shall give an exposition of the treatment of the classical theory for which time-periodic approximate solutions emerge in the weakly nonlinear regime. This material is in principle completely standard and discussed in textbooks and reviews \cite{murdock,KM}, but it is not included in most theoretical physics curricula and may be unfamiliar to the reader. (Further sense of disorientation may emerge from the variety of names applied to these techniques: they are likely to be referred to as `multi-scale' or `time-averaging' methods by applied mathematicians, and as `effective equation' and `resonant approximation' by the pure math communities working on nonlinear PDEs.) Somewhat surprisingly, once the system is quantized, the counterpart of these methods is something universally known: the Rayleigh-Schr\"odinger perturbation theory for the energy shifts in degenerate quantum spectra due to perturbations. This will be explained in section~\ref{sec:quantum}.

\subsection{Weakly nonlinear dynamics of strongly resonant systems}\label{secpertgen}

We start with a completely elementary example that nonetheless shows how naive perturbative expansions become inadequate to describe long-term dynamics of weakly nonlinear classical systems, and offers a remedy. Consider a weakly anharmonic oscillator with the Hamiltonian
\beq
H=\frac{p^2+\omega^2 x^2}{2}+\frac{gx^4}4,
\eeq
where $g$ is treated as a small parameter. The equation of motion is
\beq
\ddot x+\om^2 x+gx^3=0.
\eeq
When $g=0$,
\beq
x(t)\equiv x_0(t)=\frac{1}{\sqrt{2\om}}\left(\al e^{-i\om t} +\ab e^{i\om t}\right),
\label{hox0}
\eeq
where $\al$ is a complex constant and the overall normalization is chosen for future convenience. One could naively try to perturbatively improve this solution at order $g$ by writing $x(t)= x_0(t)+g x_1(t)$ where $x_1(t)$ satisfies
\beq
\ddot x_1+\om^2 x_1+x_0^3=0.
\eeq
This is solved by
\beq
x_1=\frac1{\sqrt{2\om}}\left(\frac{3ie^{i\om t}}{8\om^3}(i+2t\om)\al\ab^2+\frac{e^{3i\om t}}{16\om^3}\ab^3+\mathrm{c.c.} \right)
\label{hox1sol}
\eeq
(up to an arbitrary solution of $\ddot x_1+\om^2 x_1=0$, which simply fixes how the initial conditions depend on $g$). The problem with the above solution is that the first term grows with $t$ and, at $t\sim 1/g$, starts competing with $x_0$ in magnitude, invalidating the naive perturbation theory. Such terms are known as {\it secular terms}. Naive perturbation theory of course remains valid for predictions on $g$-independent time scales, where it gives tiny corrections of order $O(g)$ to (\ref{hox0}). Such tiny corrections are, however, rarely of interest. What is interesting is to track the leading effect of small nonlinearities over long times where they give corrections of order 1 to the unperturbed solution. This goal requires alternative approaches to the perturbative treatment.

There are many ways to construct an improved perturbative treatment that captures the regime of interest. We shall focus on the {\it time-averaging} approach that goes back to the classic works of Bogoliubov and Krylov almost a hundred years ago. To this end, we perform a canonical transformation from $(x,p)$ to a pair of conjugate complex variables $(\al,\ab)$ patterned on the unperturbed solution (\ref{hox0}):
\beq
x(t)=\frac{1}{\sqrt{2\om}}\left(\al(t) e^{-i\om t} +\ab(t) e^{i\om t}\right),\qquad p(t)=i\sqrt{\frac{\om}2}\left(\ab(t) e^{i\om t}-\al(t) e^{-i\om t} \right).
\label{hoint}
\eeq
(This transformation could be habitually called going to the interaction picture by a quantum field theorist.) The resulting equation of motion for $\al$ reads
\begin{align}
\dot\al\equiv \frac{d}{dt}\left[\left(\sqrt{\frac{\om}2}x+i\frac1{\sqrt{2\om}}p\right)e^{i\om t}\right] &=-\frac{ig}{4\om^2}\left(\al e^{-i\om t} +\ab e^{i\om t}\right)^3e^{i\om t}\label{eomho}\\
&=-\frac{ig}{4\om^2}(3\ab\al^2+3\ab^2\al e^{2i\om t}+\al^3 e^{-2i\om t}+\ab^3e^{4\om t}).\nonumber
\end{align}
We can see that the derivative of $\al$ is of order $g$ so that $\al$ varies slowly, on time scales of order $1/g$. At the same time, all terms on the right-hand side oscillate with periods of order 1, except for the first term. The idea of time-averaging is that all such oscillating terms `average out' at leading order and can be neglected. This can be proved rigorously \cite{murdock}. The remaining averaged equation is
\beq
\dot\al=-\frac{3ig}{4\om^2}\ab\al^2,
\label{resho}
\eeq
which is simply solved by
\beq
\al(t)=\al(0) \exp\left[-\frac{3ig}{4\om^2}|\al(0)|^2t\right].
\label{hoalsol}
\eeq
Substituting this back into (\ref{hoint}), we see that the leading effect of nonlinearities is to simply shift the oscillation frequency $\om$ by a small amplitude-dependent correction $3g|\al(0)|^2/4\om^2$, known as the Poincar\'e-Lindstedt shift. Evidently, naively expanding (\ref{hoalsol}) in powers of $g$ would have recovered the secular term in (\ref{hox1sol}), but that would have damaged the uniform applicability of our solution on time scales of order $1/g$.

We now turn to a system of coupled oscillators with frequencies $\om_n$, which are a prototype for all classical field systems. Consider the Hamiltonian
\beq
H=\sum_n \frac{p_n^2+\om_n^2 x_n^2}2 \,\,+ \,\,\frac{g}{4} \sum_{nmkl} S_{nmkl} x_n x_m x_k x_l,
\label{hammultosc}
\eeq
where $S_{nmkl}$ is fully symmetric under interchanges of $n,m,k,l$.
The idea behind its treatment at small $g$ is similar to the trivial example above, but much richer structures may emerge depending on the spectrum of $\om_n$. At $g=0$, the system consists of independent oscillators. Attempting to construct naive perturbation theory around the corresponding solutions results in secular terms that invalidate the perturbative expansion on time scales of order $1/g$. Instead, one goes to the interaction picture, as in (\ref{hoint}), by writing
\beq
x_n(t)=\frac{1}{\sqrt{2\om_n}}\left(\al_n(t) e^{-i\om_n t} +\ab_n(t) e^{i\om_n t}\right),\qquad p_n(t)=i\sqrt{\frac{\om_n}2}\left(\ab_n(t) e^{i\om_n t} -\al_n(t) e^{-i\om_n t}\right).
\label{intpixp}
\eeq
As an equation of motion for $\al_n$, one gets the following generalization of (\ref{eomho}):
\bea
&\dsty\dot \al_n =-i g \sum_{mkl} \frac{S_{nmkl}}{4\sqrt{\om_n\om_m\om_k\om_l}}\left(\al_m(t) e^{-i\om_m t} +\ab_m(t) e^{i\om_m t}\right)\left(\al_k(t) e^{-i\om_k t} +\ab_k(t) e^{i\om_k t}\right)\nonumber\\
&\dsty\hspace{7cm}\times\left(\al_l(t) e^{-i\om_l t} +\ab_l(t) e^{i\om_l t}\right)e^{i\om_n t}.\label{eomhomult}
\eea
We then proceed applying time-averaging to this equation, giving a valid approximation at small $g$ on time scales $1/g$, by discarding all explicitly oscillating terms on the right-hand side. Non-oscillating terms (which may generate large effects over long time scales) are defined by the condition
\beq
\om_n\pm \om_m\pm\om_k\pm\om_l=0,
\label{genres}
\eeq
where the three plus-minus signs are independent; they correlate with whether $\ab$ or $\al$ is chosen when expanding the product of the three bracketed expressions in (\ref{eomhomult}).

The structure of the resonant approximation (i.e., the time-averaged system) obtained by keeping only terms satisfying (\ref{genres}) on the right-hand side of (\ref{eomhomult}) crucially depends on how {\it resonant} is the spectrum of normal mode frequencies $\om_n$. `Resonant' here is used in a sense typical of Hamiltonian perturbation theory and the KAM theorem \cite{arnold}, namely, a spectrum possesses a resonance if there exists a set of not simultaneously vanishing integers $n_k$ such that
\beq
\sum_k n_k\om_k=0.
\label{resspectr}
\eeq
If no relations of these form exist, the only way to satisfy (\ref{genres}) is by forming combinations $\om_n+\om_m-\om_n-\om_m$, which are trivially zero irrespectively of the form of the spectrum. Such trivial resonances, however, provide only very few non-oscillating terms on the right-hand side of (\ref{eomhomult}). Keeping only these terms results in a very simple resonant approximation of the form $\dot\al_n=i\Omega_n(|\al|^2)\al_n$, where $\Omega_n$ are linear functions of the absolute value squared of all $\alpha$'s, but not of their phases. As a result, the equations completely decouple and are integrated as $\al_n(t)=\exp[ig\,\Omega_n(|\al(0)|^2)\,t]\,\al_n(0)$, which is a direct analog of the one-dimensional Poincar\'e-Lindstedt shift (\ref{hoalsol}). The absolute values of the normal mode amplitudes $|\al_n|^2$ do not depend on time in the absense of resonances (\ref{resspectr}), and hence there is no appreciable energy transfer between the modes due to nonlinearities at small $g$. The only significant effect of the nonlinearities for a nonresonant frequency spectrum is to give small corrections to the frequencies proportional to $g$.

The situation becomes much more complicated and interesting when resonances of the form (\ref{resspectr}) are present, as will be manifested by
our subsequent studies of the AdS dynamics. In such cases, there are nontrivial solutions to (\ref{genres}). For example, if $\om_n$ is a linear function of $n$, resonances of the form $\om_n+\om_m-\om_k-\om_l=0$ are present for any $n+m=k+l$. With extra terms present in the resonant approximation to (\ref{eomhomult}) the equations for different $\al_n$ no longer decouple from each other, and the absolute values $|\al_n|^2$ are no longer constant. In physical terms, nontrivial resonances (\ref{resspectr}) enable significant energy transfer between the different modes on time scales of order $1/g$, no matter how small $g$ is. This is what makes weakly nonlinear dynamics of strongly resonant systems highly non-trivial.

The precise form of the resonant approximation to (\ref{eomhomult}) depends on the precise structure of the resonant spectrum of $\om_n$. Assuming, as in the previous paragraph, that $\om_n$ is a linear function of $n$ and the only relevant resonances (\ref{genres}) are those with $n+m=k+l$ (which will be the case in our subsequent AdS considerations), and then retaining only non-oscillatory terms on the right-hand side of (\ref{eomhomult}) results in
\beq
i\dot\al_n=3  g \sum_{n+m=k+l} C_{nmkl} \ab_m \al_k\al_l,
\eeq
where $C_{nmkl}=S_{nmkl}/(4\sqrt{\om_n\om_m\om_k\om_l})$. Then, defining the {\it slow time} $\tau\equiv 3gt$, we arrive at a {\it resonant system} of a form that will be crucial for the rest of our considerations:
\beq
i\frac{d\al_n}{d\tau}=  \sum_{n+m=k+l} C_{nmkl} \ab_m \al_k\al_l.
\label{reseq}
\eeq
Of course, to give substance to this quick sketch of weakly nonlinear resonant dynamics, it remains to show accurately how the weakly nonlinear dynamics of fields in AdS can be recast as (\ref{reseq}) and to specify the corresponding expression for the {\it interaction coefficients} $C_{nmkl}$.

\subsection{AdS perturbations}

Our goal is to develop a treatment of small amplitude nonlinear fields in {\em global} Anti-de Sitter spacetime denoted as AdS$_{d+1}$, where $d$ is the number of spatial dimensions. The AdS metric 
can be written as
\beq
ds^2=\frac{1}{\cos^2{x}}\left(-dt^2+dx^2+\sin^2{x}\,d\Omega_{d-1}^2\right),
\label{adsmetric}
\eeq
where the curvature radius has been set to 1 by a choice of length units, and
$d\Omega_{d-1}^2$ is the round metric on the $(d-1)$-sphere parametrized by hyperspherical coordinates $\Omega=\left\lbrace \theta_{1},\cdots, \theta_{d-2}, \varphi \right\rbrace$. 

As it turns out, AdS spacetimes operate as cavities with a purely discrete, highly resonant spectrum of normal mode frequencies. Any field system placed in this cavity can then be represented as an infinite
set of interacting oscillators with resonant relations among their frequencies. This is precisely the setup of (\ref{hammultosc}), and hence weakly nonlinear dynamics of fields in AdS can be treated following the general guidelines given above for coupled oscillator systems.

We shall demonstrate these ideas explicitly using a complex scalar in AdS with quartic self-interactions, which will form the basis of our technical consideration in this paper. For that, we will closely follow the discussion in \cite{BEL}. Then we shall briefly explain how things work for other field systems.
The Lagrangian for a complex scalar field $\phi$ is simply
\beq\label{Lagr}
S=\int  \left\{|\del\phi|^2+m^2|\phi|^2+\frac{\lambda |\phi|^4}2\right\} \sqrt{-g}\,\,d^{d+1} x.
\eeq
In the background (\ref{adsmetric}), the equations of motion are
\beq\label{AdSwave}
\cos^2{x}\left(-\del_t^2\phi+\frac{1}{\tan^{d-1}{x}}\del_x\left(\tan^{d-1}{x}\del_x\phi\right)+\frac{1}{\sin^2{x}}\De_{S^{d-1}}\phi\right)-m^2\phi=\lambda |\phi|^2\phi,
\eeq
where $\De_{S^{d-1}}$ is the $(d-1)$-sphere Laplacian, which can be defined recursively,
\beq
\Delta_{S^{d}}=\frac{1}{\sin^{d-1}\te_{d-1}}\del_{\te_{d-1}}\left(\sin^{d-1}\te_{d-1}\del_{\te_{d-1}}\right)+\frac{1}{\sin^2\te_{d-1}}\Delta_{S^{d-1}},
\eeq
with $\Delta_{S^{1}}=\partial^2_{\phi}$.
The linearized system, obtained from (\ref{AdSwave}) by replacing the right-hand side with zero, can be solved by separation of variables. One starts by solving the eigenvalue problem 
\beq \label{eq_eigenvalAdS}
\left(\frac{1}{\tan^{d-1}{x}}\del_x\left(\tan^{d-1}{x}\del_x\right)+\frac{1}{\sin^2{x}}\De_{S^{d-1}}-\frac{m^2}{\cos^2{x}}\right)\ei(x,\Omega)=-\oi^2\ei(x,\Omega)
\eeq
for the mode functions $\ei(x,\Omega)$ and the corresponding frequencies $\oi$.
Here, as usual for systems with spherical symmetry, the index $n$ labels the radial overtone, while $l$ and $k$ characterize the angular momentum state, as will be discussed below. One then expands the field $\phi$ in terms of these mode functions, which yields the general linearized solution
\beq\label{AdSlin}
\phi_{\mbox{\tiny linear}}(t,x,\Omega)=\sum_{n=0}^{\infty}\sum_{l,k}(A_{\indices}e^{-i\oi t}+B_{\indices}e^{i\oi t})\,\ei(x,\Omega),
\eeq
with arbitrary complex constants $A_{\indices}$ and $B_{\indices}$. 

\begin{figure}[t]
\centering
\includegraphics[scale=0.5]{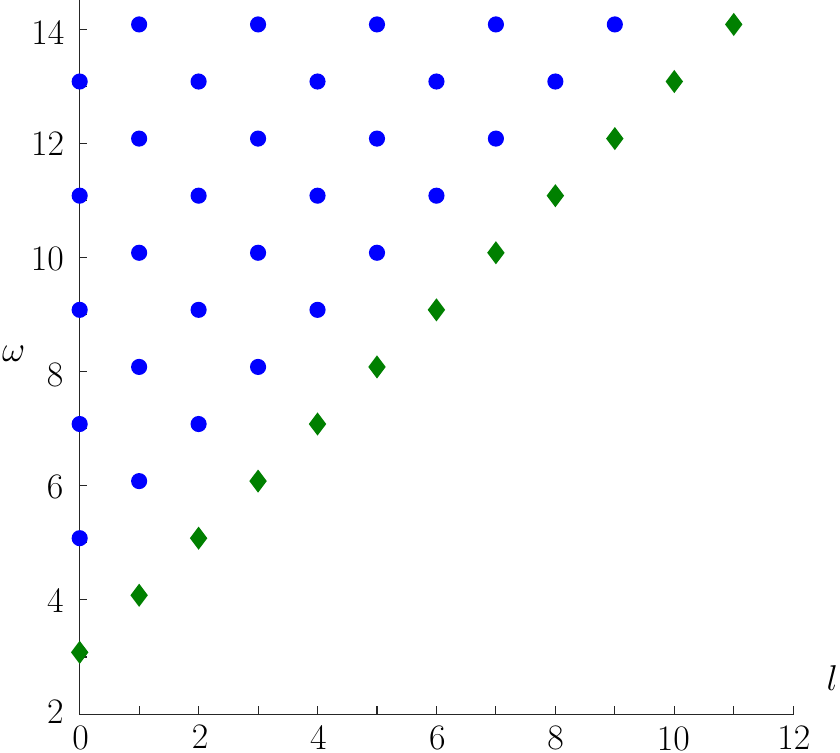}
\caption{The frequencies of the linearized normal modes \eqref{eq_eigenvalAdS} for different values of $n$ and $l$. Each dot represents a full multiplet of states with angular momentum $l$ containing $(l+d-3)!(2l+d-2)/(l!(d-2)!)$ states with degenerate frequencies given by \eqref{eq_eigenvalueAdS} and evaluated at $d=3$ and $m=1/2$. The multiplets containing the maximally rotating modes, to play a crucial role in our later considerations, are highlighted and displayed as green rhombi.}
\label{figtower}
\end{figure}

The mode functions are explicitly known and given by
\beq \label{eq_modefunctionAdS}
\ei(x,\Omega)=\mathcal{N}_{\indices}\cos^{\de}{x}\sin^l{x}P_n^{\left(\de-\frac{d}{2}, l+\frac{d}{2}-1\right)}(-\cos{2x})Y_{lk}(\Omega),
\eeq
with associated frequencies
\beq \label{eq_eigenvalueAdS}
\oi=\de+2n+l,
\eeq
where $\de=\frac{d}{2}+\sqrt{\frac{d^2}{4}+m^2}$ and $\mathcal{N}_{\indices}$ is a normalization factor. (It follows from \eqref{eq_eigenvalueAdS} that the difference of any two frequencies is integer irrespectively of $\de$.) The $P_n^{(a,b)}(x)$ are the Jacobi polynomials and form an orthogonal basis on the interval $(-1,1)$ with respect to the measure $(1-x)^a(1+x)^b$. The $Y_{lk}$ are spherical harmonics on a $(d-1)$-dimensional sphere, i.e.\ eigenfunctions of the sphere Laplacian $\Delta_{S^{d-1}}$ with eigenvalues $l(l+d-2)$. The index $k$ labels the spherical harmonics contained in a given $l$-multiplet.
The number of values $k$ takes can be deduced by remembering that the spherical harmonics
of angular momentum $l$ form a representation of $SO(d)$ that is a rank $l$ fully symmetric fully traceless
tensor. The number of independent components of such a tensor can be counted as the number of components of a rank $l$ fully symmetric tensor in $d$ dimensions, which is $(l+d-1)!/(l!(d-1)!)$,
minus the same for a similar tensor of rank $l-2$. This yields $(l+d-3)!(2l+d-2)/(l!(d-2)!)$, which is the number of values $k$ takes. 
The resonant tower of frequencies given by (\ref{eq_eigenvalueAdS}) can be visualized as in Fig.~\ref{figtower}.
The full set of modes of a given frequency (all the dots on a given horizontal line in Fig.~\ref{figtower}) has precisely the number of components of a fully symmetric tensor of rank $l$, where $l$ is the maximal angular momentum within that frequency level.
Such tensors form irreducible representations of $SU(d)\supset SO(d)$, which is a hidden symmetry
of the problem and can be made manifest by relating the eigenvalue problem (\ref{eq_eigenvalAdS})
to a particular superintegrable quantum-mechanical system on a sphere \cite{hggs1,hggs2} known as the `Higgs oscillator' \cite{Higgs,Leemon}.

To employ time-averaging we first perform a canonical transformation from $\phi$ and its conjugate momentum to the complex amplitudes $\ali(t)$ and $\bei(t)$ in a manner analogous to (\ref{intpixp}), so that $\phi$ is given by
\beq\label{AdSint}
\phi(t,x,\Omega)=\sum_{n=0}^{\infty}\sum_{l,k}\frac1{\sqrt{2\om_{nlk}}}(\al_{\indices}e^{-i\oi t}+\bar\be_{\indices}e^{i\oi t})\ei(x,\Omega).
\eeq
From this, and the corresponding formula for the momentum conjugate to $\phi$, as in (\ref{intpixp}), one gets
\beqa \label{eq_preaveragingS3alpha}
&i\dsty\dot\al_{nlk}=\lambda\hspace{-1mm}\sum_{n_1l_1k_1}\sum_{n_2l_2k_2}\sum_{n_3l_3k_3}C_{nlk,n_1l_1k_1,n_2l_2k_2,n_3l_3k_3}\,\,\bar c_{n_1l_1k_1}\,c_{n_2l_2k_2}\,c_{n_3l_3k_3}\,e^{i\om_{nlk} t},\\
\label{eq_preaveragingS3beta}
&i\dsty\dot{\bar{\be}}_{nlk}=-\lambda\hspace{-1mm}\sum_{n_1l_1k_1}\sum_{n_2l_2k_2}\sum_{n_3l_3k_3}C_{nlk,n_1l_1k_1,n_2l_2k_2,n_3l_3k_3}\,\,\bar c_{n_1l_1k_1}\,c_{n_2l_2k_2}\,c_{n_3l_3k_3}\,e^{-i\om_{nlk} t}.\rule{0mm}{8mm}
\eeqa
Here, $c_{nlk}\equiv\left(\al_{nlk} e^{-i\om_{nlk}t}+\bar\be_{nlk} e^{i\om_{nlk} t}\right)$ and
\beq
C_{nlk,n_1l_1k_1,n_2l_2k_2,n_3l_3k_3}=\int_0^{\frac{\pi}{2}}dx\frac{\tan^{d-1}{x}}{\cos^2{x}} \int d\Omega_{d-1} \,\,\frac{\dsty \bar{e}_{nlk}\bar{e}_{n_1l_1k_1}e_{n_2l_2k_2}e_{n_3l_3k_3}}{\dsty 4\sqrt{\om_{nlk}\om_{n_1l_1k_1}\om_{n_2l_2k_2}\om_{n_3l_3k_3}}}.
\label{Cgen}
\eeq
Note that angular momentum conservation imposes constraints on these interaction coefficients.
Namely, each spherical harmonic in (\ref{eq_modefunctionAdS}) labelled by $lk$ carries a definite value
of the angular momentum projection on the polar axis, which we denote $m(l,k)$.
The angular momentum conservation means that the interaction coefficient $C$ vanishes
unless $m(l,k)+m(l_1,k_1)=m(l_2,k_2)+m(l_3,k_3)$. We will schematically refer to this condition below as $m + m_1 = m_2+m_3$.

The time-averaging then proceeds as per general guidelines discussed in section \ref{secpertgen}. The  amplitudes $\al$ and $\be$ vary slowly, so that significant changes may only occur on time scales of order $1/\lambda$. On these time scales, terms on the right-hand side containing explicit oscillatory functions of time can never contribute significantly, while the effect of terms without explicit oscillations may build up to contributions of order 1 to $\al$ and $\be$. The time averaging (or resonant approximation) then amounts to keeping only these latter terms.   

Due to the highly resonant spectrum of AdS normal mode frequencies given by (\ref{eq_eigenvalueAdS}), there are many ways for the oscillatory factors within the individual terms on the right-hand side of (\ref{eq_preaveragingS3alpha}-\ref{eq_preaveragingS3beta}) to cancel each other. For generic values of $\de$, this occurs whenever $\om_{nlk}+\om_{n_1l_1k_1}= \om_{n_2l_2k_2}+\om_{n_3l_3k_3}$ (or other similar relations obtained by permuting the groups of indices 1, 2 and 3). This condition translates to simply $2n+l+2n_1+l_1=2n_2+l_2+2n_3+l_3$ (note that $\de$ cancels out in this formula). If $\de$ is an integer, as it is for massless fields, even more possibilities to satisfy the resonant condition (\ref{genres}) exist a priori. A somewhat surprising result is that these resonances do not contribute since the corresponding mode couplings (\ref{Cgen}) vanish. A derivation of these selection rules and further discussion can be found in \cite{Yang,EN}.

As a result of the selection rules for the interaction coefficients, irrespectively of the value of $\de$, the resonant approximation to (\ref{eq_preaveragingS3alpha}-\ref{eq_preaveragingS3beta}) takes the form
\beqa \label{resalgen}
&i\dsty\frac{d\al_{nlk}}{d\tau}=&\hspace{-3mm}\sum_{\substack{\om+\om_2=\om_1+\om_3 \\ m +m_1 = m_2+ m_3}}\hspace{-3mm}C_{nlk,n_1l_1k_1,n_2l_2k_2,n_3l_3k_3}\,\,\bar \al_{n_1l_1k_1}\,\al_{n_2l_2k_2}\,\al_{n_3l_3k_3}\\
&&\hspace{2cm}+2\hspace{-3mm}\sum_{\substack{\om+\om_2=\om_1+\om_3  \\ m+m_1 = m_2+ m_3}}\hspace{-3mm}C_{nlk,n_1l_1k_1,n_2l_2k_2,n_3l_3k_3}\,\,\bar \be_{n_1l_1k_1}\,\be_{n_2l_2k_2}\,\al_{n_3l_3k_3},\nonumber\\
\label{resbegen}
&i\dsty\frac{d\be_{nlk}}{d\tau}=&\hspace{-3mm}\sum_{\substack{\om+\om_2=\om_1+\om_3  \\ m+m_1 = m_2+ m_3}}\hspace{-3mm}C_{nlk,n_1l_1k_1,n_2l_2k_2,n_3l_3k_3}\,\,\bar \be_{n_1l_1k_1}\,\be_{n_2l_2k_2}\,\be_{n_3l_3k_3}\\
&&\hspace{2cm}+2\hspace{-3mm}\sum_{\substack{\om+\om_2=\om_1+\om_3  \\ m+m_1 = m_2+ m_3}}\hspace{-3mm}C_{nlk,n_1l_1k_1,n_2l_2k_2,n_3l_3k_3}\,\,\bar \al_{n_1l_1k_1}\,\al_{n_2l_2k_2}\,\be_{n_3l_3k_3},\rule{0mm}{8mm}\nonumber
\eeqa
where we have introduced the slow time $\tau=\lambda t$, and the following shorthands: $\om=\om_{nlk},\om_i=\om_{n_il_ik_i}$. We have also displayed explicitly the angular momentum conservation constraint in the sums, as clarified under (\ref{Cgen}).

Note that symmetries got enhanced in transition from (\ref{eq_preaveragingS3alpha}-\ref{eq_preaveragingS3beta}) to (\ref{resalgen}-\ref{resbegen}). Namely, while  (\ref{eq_preaveragingS3alpha}-\ref{eq_preaveragingS3beta}) is only invariant with respect to rotating all $\al_{nlk}$ by a common phase and all $\be_{nlk}$ by the opposite phase,  (\ref{resalgen}-\ref{resbegen}) allows rotating all $\al_{nlk}$ by a common phase and all $\be_{nlk}$ by a completely independent common phase. Thus, the usual charge $U(1)$ of the complex scalar got enhanced to $U(1)\times U(1)$. One can think of it pictorially as the number of particles and antiparticles being conserved independently
(within the resonant approximation, and hence valid at leading order on time scales $1/\lambda$). This will, of course, have pronounced consequences when we turn to the quantum theory. Within the classical equations (\ref{resalgen}-\ref{resbegen}), the symmetry enhancement implies, in particular, that one can consistently set $\be$ to zero. We shall henceforth focus on this simple truncation given by
\beq
i\dsty\frac{d\al_{nlk}}{d\tau}=\hspace{-3mm}\sum_{\substack{\om+\om_2=\om_1+\om_3  \\ m+m_1 = m_2+ m_3}}\hspace{-3mm}C_{nlk,n_1l_1k_1,n_2l_2k_2,n_3l_3k_3}\,\,\bar \al_{n_1l_1k_1}\,\al_{n_2l_2k_2}\,\al_{n_3l_3k_3}.
\label{alrestrunc}
\eeq

\subsection{The maximally rotating sector}
\label{secmxrota}

With a huge tower of AdS normal modes, equations (\ref{alrestrunc}) remain rather unmanageable.
They do admit, however, consistent truncations to many smaller sets of modes, and these truncations
may often be more tractable.

One particular truncation of this type that we shall focus on restricts the normal modes to those
with the maximal amount of angular momentum for a given frequency. Indeed, there is one specific mode of any given frequency that has the largest value
of the angular momentum projection on the polar axis. From every frequency level, one then chooses precisely this mode (the angular momentum multiplets from which these modes originate are highlighted in Fig.~\ref{figtower}). These modes are directly analogous to what is known under the name of the `lowest Landau level' in the literature on trapped Bose-Einstein condensates \cite{BBCE,BBCE2}.

To see that the truncation is consistent, consider (\ref{alrestrunc}) and imagine that only maximally
rotating modes are initially turned on. In this case, the only nonzero contributions are from $n_1=n_2=n_3=0$ and hence $\om_1=\de+l_1$, $\om_2=\de+l_2$, $\om_3=\de+l_3$. Hence, $\om=\de+l_2+l_3-l_1$ and 
\beq
2n+l=l_2+l_3-l_1.
\label{nl123}
\eeq 
At the same time, the angular momentum is conserved, which is encoded in the mode couplings $C$. With respect to the polar axis used for defining maximal rotation, the $n_1l_1k_1$ mode has $l_1$ units in its angular momentum projection, and similarly $l_2$ for the $n_2l_2k_2$ mode, and $l_3$ for $n_3l_3k_3$. Angular momentum projections simply add up (and change sign under complex conjugation of the mode function), so
the given contribution on the right-hand side of (\ref{alrestrunc}) is only nonzero if mode $nlk$ has the angular momentum projection given by $l_2+l_3-l_1$. But such a state can only be in the angular momentum multiplet if $l\ge l_2+l_3-l_1$, which together with (\ref{nl123}) implies $n=0$ and $l=l_2+l_3-l_1$. Hence, mode $nlk$ precisely matches the definition of a maximally rotating mode, and no other modes, besides the maximally rotating ones, will ever get excited by equation (\ref{alrestrunc}) if none of them are excited in the initial state.

Looking at the explicit expressions \crefrange{eq_modefunctionAdS}{eq_eigenvalueAdS}, to extract
the maximally rotating modes, one must evidently choose $n=0$ as increasing $n$ just increases the frequency without changing the angular momentum. Then, from the corresponding angular momentum multiplet labelled by $k$, one must choose the spherical harmonic that has the biggest projection of the angular momentum (equal to $l$) on the polar axis. An explicit expression for such mode functions, which are labelled by one index, is
\beq
e_l(x,\te_1,\dots, \te_{d-2},\ph)=\sqrt{\frac{\G{l+1+\de}}{\pi^{d/2} \G{1+\de-\frac{d}{2}} \G{l+1}}}\cos^{\de}{x} \sin^l{x} \sin^l{\te_1}\dots\sin^l{\te_{d-2}}e^{-il\ph},
\eeq
where we have written out explicitly the hyperspherical angles $\te_1,\ldots\te_{d-2},\phi$ comprising $\Omega$.
The interaction coefficients are then evaluated from (\ref{Cgen}) as
\beq
C_{nmjk}= 
 \frac{\G{2\de - \frac{d}{2}} }{4 \pi^{d/2} \G{\de - \frac{d}{2}+1}^2}\sqrt{\frac{\G{n+\de}\G{m+\de}\G{j+\de}\G{k+\de}}{\G{n+1}\G{m+1}\G{j+1}\G{k+1}}}\frac{\G{n+m+1}}{\G{n+m+2\de}}.
\label{Cmaxrt}
\eeq
Any numerical coefficient multiplying the interaction coefficient is irrelevant and can be absorbed in a redefinition of the slow time $\tau$ by redefining the coupling $\lambda$. In the following, we will therefore use the convention $C_{0000}=1$, as in \cite{solvable}, so that the interaction coefficients become
\beq
C_{nmjk}= 
 \frac{\G{2\de} }{\G{\de}^2}\sqrt{\frac{\G{n+\de}\G{m+\de}\G{j+\de}\G{k+\de}}{\G{n+1}\G{m+1}\G{j+1}\G{k+1}}}\frac{\G{n+m+1}}{\G{n+m+2\de}}.
\label{Cmaxrt}
\eeq
Thereafter, one arrives at the equation (\ref{alrestrunc}) consistently truncated to the maximally rotating modes in the form
\beq
i\frac{d\al_n}{d\tau}=\sum_{m=0}^{\infty}\sum_{k=0}^{n+m}C_{nmk,n+m-k}\alb_m\al_k\al_{n+m-k}.
\label{maxrtres}
\eeq
This  equation could be derived from the `resonant' Hamiltonian
\beq
H_{res}=\frac12 \sum_{n+m=k+l} C_{nmkl}\,\ab_n\ab_m \al_k\al_l,
\label{Hresclas}
\eeq
assuming that the canonical momentum conjugate to $\al_n$ is $i\ab_n$. It possesses two obvious conservation laws
\beq\label{AdSconserved}
N=\sum_{n=0}^{\infty}|\alpha_n|^2,\qquad
M=\sum_{n=0}^{\infty}n|\alpha_n|^2,
\eeq
which can be thought of as the conservation of the `particle number' and the total energy of the linearized theory, respectively. Conservation of $N$ relies crucially on the selection rules for the interaction coefficients and the symmetry enhancement in the resonant approximation mentioned above (\ref{alrestrunc}).

The two conservation laws mentioned above are generic for equations of the form (\ref{maxrtres}), irrespectively of the expression for the interaction coefficients $C$. In (\ref{maxrtres}) with the specific interaction coefficients (\ref{Cmaxrt}), one has, however, an extra conserved quantity
\beq
Z=\sum_{n=0}^{\infty}\sqrt{(n+1)(n+\de)}\,\ab_{n+1}\al_n.
\label{Zclas}
\eeq
Conservation of $Z$ can be traced back \cite{breathing} to the fact that the center-of-mass of any system in AdS performs
simple perfectly periodic motions irrespectively of the complexity of the dynamics of other degrees of freedom (just as the center-of-mass in Minkowski space moves with a constant velocity). Conservation of $Z$ hints at solvable features in (\ref{maxrtres}) that we shall explore below, while the fact that it comes from something as generic as the center-of-mass motion in AdS makes one expect that similar features would be seen in other, more complicated systems, and not just for the self-interacting probe scalar (\ref{Lagr}). While the center-of-mass motion in AdS may seem a triviality, it imposes powerful relations between the mode couplings in the normal mode basis \cite{breathing} and leads to solvable features in the corresponding resonant systems.


\subsection{Time-periodicities}
\label{sectimeper}

A general theory of resonant systems of the form (\ref{maxrtres}) admitting a conservation law of the form
(\ref{Zclas}) was developed in \cite{solvable}. It turns out that the relations between the interaction coefficients $C$ this conservation law implies are also responsible for the existence of simple dynamically invariant manifolds where the evolution can be analyzed exactly. This evolution, furthermore, turns out to display time-periodic features which form the core of our current study. We demonstrate these features below using an adapted formulation of derivations from \cite{BEL} where they were originally described for the maximally rotating sector of a complex scalar in AdS.

We start with the following ansatz 
\beq \label{eq_adsansatz}
\al_n(\tau) \equiv f_n \beta_n(\tau) =f_n(b(\tau)+na(\tau))\,(p(\tau))^n, 
\eeq
where $f_n$ is a set of numbers given by 
\beq \label{fn}
f_n = \sqrt{\frac{(\de)_n}{n!}}.
\eeq
Here, $(x)_n \equiv {\G{x+n}}/{\G{x}}$ is the rising Pochhammer symbol. We then demonstrate that the resonant system (\ref{maxrtres}) respects the proposed ansatz. (Note that $\beta_n$ is just a rescaling of $\alpha_n$, and unrelated to the amplitudes $\beta_{nlk}$ appearing above, which we have set to zero.)

The evolution equations \eqref{maxrtres} written in terms of $\beta_n$ become
\beq \label{eq_adsflow}
i\frac{d\be_n}{d\tau}= \frac{\G{2\de}}{\G{\de}^3}\sum_{m=0}^{\infty}\frac{\G{m+\de}}{\G{m+1}}\sum_{k=0}^{n+m}\binom{n+m}{k}\B{k+\de}{n+m-k+\de}\beb_m\be_k\be_{n+m-k},
\eeq
using the beta function $B(x,y) \equiv {\G{x}\G{y}}/{\G{x+y}}$. Inserting the ansatz \eqref{eq_adsansatz} in \eqref{eq_adsflow}, one can compute the various sums over $k$ using the integral representation of the beta function as follows
\beq
\begin{aligned}
\sum_{k=0}^{N}\binom{N}{k}\B{k+\de}{N-k+\de}&=\int_0^1dx\sum_{k=0}^N\binom{N}{k}x^{k+\de-1}(1-x)^{N-k+\de-1}\\
&=\int_0^1dx\ x^{\de-1}(1-x)^{\de-1}=\B{\de}{\de}, \\
\sum_{k=0}^{N}k\binom{N}{k}\B{k+\de}{N-k+\de}&=N\B{\de+1}{\de}=\frac{N}{2}\B{\de}{\de},\\
\sum_{k=0}^{N}k^2\binom{N}{k}\B{k+\de}{N-k+\de}
&=\left(\frac{N}{2}+\frac{N(N-1)}{2}\frac{1+\de}{1+2\de}\right)\B{\de}{\de}.
\end{aligned}
\eeq
The $m$-summation can afterwards be carried out using
\beq
\sum_{m=0}^{\infty}\frac{\G{m+\de}}{\G{m+1}}m^Ax^m=(x\del_x)^A\frac{\G{\de}}{(1-x)^{\de}}.
\eeq
Performing these summations in \eqref{eq_adsflow} results in an equation with quadratic polynomials in $n$ on both sides, which shows the consistency of the ansatz \eqref{eq_adsansatz} and provides three equations of motion for $a(\tau)$, $b(\tau)$ and $p(\tau)$:
\begin{align}
\frac{i\dot{b}}{(y+1)^{\de}}&=b|b|^2+y\de\left(\bar{a} b^2+a|b|^2+|a|^2b\right)+ \nonumber \\
& \quad y^2\de\left(|a|^2b(\de+1)+\frac{a^2\bb}{2}\frac{\de(\de+1)}{1+2\de}+a|a|^2\frac{\de(\de+1)}{1+2\de}\right)+y^3\frac{a|a|^2}{2}\frac{\de^2(\de+1)(\de+2)}{1+2\de}, \label{eq_bdot_AdS} \\
\frac{i\dot{a}}{(y+1)^{\de}}&=\frac{a|b|^2}{2}\frac{2+3\de}{1+2\de}-\frac{a^2\bb}{2}\frac{\de}{1+2\de}+y\de\left(\frac{|a|^2b}{2}\frac{2+3\de}{1+2\de}+a^2\bb\frac{\de}{1+2\de}+\frac{a|a|^2}{2}\frac{\de}{1+2\de}\right)+\nonumber\\
& \quad y^2a|a|^2\frac{\de^2(\de+1)}{1+2\de},\label{eq_adot_AdS}\\
\frac{i\dot{p}}{(y+1)^{\de}}&=\frac{p}{2}\de\left(a\bb\frac{1}{1+2\de}+y|a|^2\frac{\de}{1+2\de}\right),\label{eq_pdot_AdS}
\end{align}
where we have defined 
$$y \equiv \frac{|p|^2}{1-|p|^2}\,.$$
The equation of motion for $y$ can be derived from \eqref{eq_pdot_AdS}, giving
\beq \label{eq_ydot_AdS}
\frac{\yd}{(y+1)^{\de+1}}=\frac{y\de}{1+2\de}\Im{(\bar{b} a)}.
\eeq
The conserved quantities \eqref{AdSconserved} and \eqref{Zclas} can be expressed within our ansatz as 
\begin{align}\label{eq: conserved quantities ansatz}
N&=(y+1)^{\de}\left[|b|^2+2\Re{(a\bb)}\de y+|a|^2\de y(1+(\de+1)y)\right],\\
M&=\de y (y+1)^{\de}\left[|b|^2+2\Re{(a\bb)}(1+(\de+1)y) +|a|^2 (1 + y (1 + \de) (3 + y (2 + \de)))  \right], \\
Z&= \bar{p} \de (y+1)^{\de+1}\left[|b|^2+b\bar{a}(1+(1+\de)y)+a\bar{b}(1+\de)y + |a|^2 y (1+\de)(2+y(2+\de))\right],
\label{eq: conserved quantities ansatz 2}
\end{align}
and an additional conserved quantity $S$ follows from \eqref{eq_ydot_AdS} and the time derivative of $|a|^2$ obtained from \eqref{eq_adot_AdS}:
\beq
S=|a|^2 \de y(y+1)^{\de+1},
\eeq
which can be related to the Hamiltonian and the particle number through 
\beq
S^2 = \frac{1+2\de}{1+\de}\left(N^2 - 2H_{res}\right).
\label{eq: classical eq for energy ladder1}
\eeq
Note that $S$ can also be written in terms of $N$, $M$ and $Z$, as expressed by \crefrange{eq: conserved quantities ansatz}{eq: conserved quantities ansatz 2},
\beq
S^2 = \frac{1}{1+\de}\left(N M \de+M^2-|Z|^2\right).
\label{eq: classical eq for energy ladder2}
\eeq
By expressing $|b|^2$, $|a|^2$ and $\Re{(a\bb)}$ through $N$, $M$, $S$ and $y$ 
\begin{align}
|a|^2&=\frac{S}{y \de (y+1)^{\de+1}}, \label{adsa2}\\
\Re{(a\bb)}&=\frac{M}{2\de y(y+1)^{\de+1}}-\frac{N}{2 (y+1)^{\de+1}}-\frac{1+2(\de+1)y}{2y \de (y+1)^{\de+1}}S,\\
|b|^2&=-\frac{M}{(y+1)^{\de+1}}+\frac{1+(\de+1)y}{(y+1)^{\de+1}} N +\frac{(\de+1)y}{(y+1)^{\de+1}}S,\label{adsb2}
\end{align}
and using $\left(\textrm{Im}(\bar{a}b)\right)^2=|a|^2 |b|^2 - \left(\textrm{Re}(\bar{a}b)\right)^2$, the equation of motion for $y$ can be turned into an equation expressing the energy conservation for an ordinary one-dimensional harmonic oscillator,
\beq
\begin{split}
\yd^2=&-y^2\frac{N^2\de^2 +4(\de+1)S^2}{4(1+2\de)^2} +y\frac{M(N\de +2 S)+S(N\de-2(1+\de)S)}{2(1+2\de)^2}-\frac{(M-S)^2}{4(1+2\de)^2},
\label{eq: evolution y'2}
\end{split}
\eeq
which demonstrates that $y$, and hence $|p|$, are exactly periodic functions in the three-dimensional invariant manifold, with period
\beq \label{eq: period returns}
T=\frac{4\pi(1+2\de)}{\sqrt{N^2 \de^2+4(\de+1)S^2}}.
\eeq
From equations (\ref{adsa2}-\ref{adsb2}), one concludes that this exactly periodic behavior is shared by $|a|^2$, $|b|^2$ and $\Re(a\bar b)$, and therefore also by the absolute value of the amplitudes $|\alpha_n|$, which exhibit exact periodic returns to the initial configuration with period \eqref{eq: period returns}.

We end this section by briefly discussing the solutions of (\ref{maxrtres}) on the submanifold $Z=0$, as these solutions will be of relevance for comparing to the quantum results. In fact, the evolution on this submanifold is very simple. Indeed, the periodic returns with period \eqref{eq: period returns} that were seen in $y$, $|a|^2$, $|b|^2$ and $\Re(a\bar b)$ are not manifest since these quantities do not depend on time at all. This can be understood by first noting from \eqref{eq: conserved quantities ansatz 2} that the following relation holds in general:
\begin{equation}
\Im{pZ}=\Im{(b\bar{a})} y \de(y+1)^{\de}.
\end{equation}
From \eqref{eq_ydot_AdS} it follows that when $Z=0$, $y$ is constant. Imposing that the right-hand side of \eqref{eq: evolution y'2} should vanish, one finds that $y$ is equal to 
\begin{equation}
y=\frac{S-M}{2M+N\de},
\label{eq: y on Z is 0 manifold}
\end{equation}
and therefore
\begin{equation}
|p|^2 =\frac{M\left(\gamma-1\right)}{M\gamma+M+N\de},
\label{eq: p2 on Z is 0 manifold}
\end{equation}
where we defined $\gamma \equiv \sqrt{\frac{M+N\de}{M\left(1+\de\right)}}$ to make the formulas more compact.
Using \crefrange{adsa2}{adsb2}, it is then easily seen that 
\begin{align}
|a|^2&=\frac{\gamma(2 M+  N \de)^{2+\de}}{\delta  \left(\gamma-1\right) \left(M\gamma+M+  N \de\right)^{\de+1}}, \label{eq: a2 on Z0}\\
\Re{(a\bb)}&=\frac{(M-N) \gamma (2 M+  N \de)^{1+\de}}{\left(\gamma-1\right)
   \left(M\gamma+M+  N\de\right)^{\de+1}},\\
|b|^2&=-\frac{(M-N) \left(M(\delta +1) \gamma+M+  N\de\right)(2 M+\delta  N)^{\de}}{\left(M\gamma+M+ 
   N \de\right)^{\de+1}}.\label{eq: b2 on Z0}
\end{align}
The motions of $a$, $b$ and $p$ are then pure phase rotations. By plugging the above constant values in the equations \crefrange{eq_bdot_AdS}{eq_pdot_AdS}, one finds that  $a$ and $b$ share a common period
\begin{equation}
T_{ab} = \frac{4\pi(1+2\de)}{2N(1+2\de)-M\de}
\label{eq: period a b on Z0}
\end{equation}
while $p$ evolves periodically with period \eqref{eq: period returns} evaluated at $Z=0$.


\subsection{Similar behaviors for other systems}

We have exposed above how time-periodic behaviors on time scales $1/\lambda$ emerge in the maximally rotating sector of a complex scalar field in AdS. It is important to keep in mind, however, that the algebraic pattern underlying the derivations of the time-periodic solutions is rather general and is present in many other resonant systems. Indeed, in \cite{solvable}, an enormous family of resonant systems of the form (\ref{Hresclas}) was constructed that shares the same features: there is a three-dimensional invariant manifold of the resonant evolution where the energy spectrum of the normal modes given by $|\alpha_n|^2$ is exactly periodic in time. This family is parametrized by an arbitrary function (through which the generating function of $C_{nmkl}$ is expressed) and thus the set of resonant Hamiltonians with such time-periodic behaviors forms an infinite-dimensional space. Our considerations of the maximally rotating truncation of the resonant system for a complex scalar field in AdS are simply a special case of the structure described in \cite{solvable}.

One might wonder how the systems in the class constructed in \cite{solvable}, with their time-periodic behaviors, could arise as resonant approximations to Hamiltonian PDEs of physical interest. This question has been addressed in \cite{breathing}. The crucial ingredient for the construction of the time-periodic solutions of the resonant system \cite{solvable} is conservation of the quantity $Z$ defined by (\ref{Zclas}). This conservation law imposes a relation between the interaction coefficients $C$, which in turn guarantees the existence of the time-periodic solutions \cite{solvable}. Conservation laws of the form (\ref{Zclas}) 
naturally arise in resonant systems originating from PDEs with {\it breathing modes} \cite{breathing}. Breathing modes are functions on the phase space that evolve exactly periodically for all solutions of the equations of motion (note, importantly, that these periodicities are on time scales of order 1, not $1/\lambda$). Within the resonant approximation, such breathing modes translate \cite{breathing} into
conservation laws of the resonant system (\ref{reseq}) of the form (\ref{Zclas}). The origin of (\ref{Zclas}) for the above example of a scalar field in AdS can be thus traced back to the center-of-mass motion in AdS,
which simply follows geodesics (irrespectively of how complex the detailed evolution is) and thus provides
breathing modes. It may seem paradoxical that a triviality like the free motion of the center-of-mass
constrains the properties of the weakly nonlinear evolution captured by the resonant system. The point is
that the exactly periodic motion of the center-of-mass implies relations between the mode couplings given by (\ref{Cgen}), and these relations, in turn, translate into existence of special solutions of the nonlinear resonant equations. Note that the construction is specific to quartic nonlinearities, and there is no obvious way to generalize it to higher nonlinearities \cite{quintic}. On the other hand, quartic nonlinearities are very
generic, and it makes our construction robust. We shall see below, and it is indeed the main point of our treatment, that the implications of these special structures in resonant systems in the corresponding quantum theory are even more far-reaching.

In what other setups, in particular those of interest for AdS/CFT dualities, can one expect the structures we have described to appear? The emergence of resonant systems with periodic behaviors in the general class of systems introduced in \cite{solvable} relies on two essential ingredients. First, the original dynamical PDEs must admit breathing
modes that give rise to conserved quantities, like the one in (\ref{Zclas}), within the resonant approximation. Second, there must exist a consistent truncation of the resonant dynamics to a subset of modes labelled by
one nonnegative integer, so that the conservation law within this truncation becomes exactly (\ref{Zclas}).
The first condition is easy to satisfy in AdS, as the center-of-mass motion provides breathing modes.
Of course, some subtleties may arise when the gravitational interactions are included, but one expects
that in an asymptotically AdS space the center-of-mass motion may still be defined and will follow simple geodesics (which are all exactly periodic). For special cases, there may be further breathing modes: for example, a conformally coupled scalar admits a radial breathing mode \cite{breathing} which underlies
the time-periodic resonant dynamics described in \cite{CF}. Regarding the truncations, the maximally symmetric truncation, like the one used above for $\phi^4$ theory, would exist for any nonlinear wave equation for a complex scalar field in AdS.

From this perspective, we find the setup of AdS$_3$ particularly promising. In this setting, gravity does not have any propagating degrees of freedom and can be integrated out, leaving an effective theory for the matter motion, for example, a nonlinear wave equation for a complex scalar field. The same should be true for the more complicated higher spin theories in AdS$_3$ that likewise do not have propagating degrees of freedom, as long as they can be consistently coupled to complex scalar field matter. Such formulations have received a lot of attention in the AdS/CFT literature \cite{hspin,hspin2}, and considerable understanding has been
developed on the CFT side. Our setup is essentially guaranteed to work in this case, though technical details need to be checked and explicit computations will be much more complicated than in our current $\phi^4$ example, on account of the very complicated nonlinearities involved. In higher-dimensional AdS, truncation to the maximally rotating sector is possible, and resonant conservation laws inherited from the center-of-mass motion must exist, but {\it a priori} there are gravitational waves that couple to the matter and one is not guaranteed to end up with a simple resonant system of the form (\ref{reseq}). There are other more distantly related situations where the pattern we discuss may arise approximately, rather than exactly, and we shall comment on that in the discussion section.

As we have remarked, treating resonant gravitational dynamics in AdS outside spherical symmetry is technically very challenging, and that is  the main reason why we are focusing on a non-gravitating scalar
here to illustrate our ideas. Extensive studies of secular terms in perturbation theory for AdS gravity outside spherical symmetry have been undertaken, see for instance \cite{nonspher1,nonspher2,nonspher3,nonspher4,highpert,nonspher5}, but no explicit formulas for the interaction coefficients as functions of the mode numbers have appeared thus far (analogous formulas in spherical symmetry are available in appendix A of \cite{CEV2}). Since our treatment below relies on analytic patterns in the mode number dependence of the interaction coefficients, further developments in analytic treatments of AdS gravity outside spherical symmetry will be necessary before we can approach that case from the point of view of our current work. Some considerations for optimizing the handling of nonlinearities in this setting have appeared in \cite{highpert}, and they may turn out useful for deriving explicit formulas for the interaction coefficients. The relatively simple case of AdS$_3$ would in fact form a natural starting point for this program.


\section{Quantum energy shifts in AdS}\label{sec:quantum}

Given the treatment of classical AdS perturbations in the previous section and the relative ease of connecting the corresponding quantum theory
to the CFT side of holographic dualities, it is worth asking what serves as the quantum counterpart of the resonant perturbation theory of the previous section. It will turn out that this quantum counterpart is considerably more elementary and familiar than the classical story.

A crucial property defining the resonant approximation is that it approximates weakly nonlinear solutions at leading order $O(1)$ but in a way that
is uniformly valid on very long time scales of order $O(1/\lambda)$ when the coupling $\lambda$ is small. It is worth contrasting this setup with the naive perturbation theory (expanding classical solutions in powers of $\lambda$) that captures solutions much more accurately with precision $O(\lambda^n)$ for any desired $n$, but only on short time scales $O(1)$. On such short time scales, the effects of small nonlinearities are always small, while the resonant approximations aims at something completely different: large effects of small nonlinearities at long time scales. What approximation in
the quantum theory shares these qualities?

The most general solution of a quantum theory can be written symbolically as a familiar sum over the Hamiltonian eigenstates $|\Psi_n\rangle$:
\beq
|\Psi(t)\rangle=\sum_n  e^{-iE_n t} \langle \Psi_n|\Psi(0)\rangle\,|\Psi_n\rangle.
\label{Psieig1}
\eeq
For a system whose Hamiltonian is of the form $H=H_0+\lambda H_{int}$, the Rayleigh-Schr\"odinger perturbation theory instructs us to expand
eigenvalues and eigenvectors as power series in $\lambda$: $E_n=E_n^{(0)}+\lambda E_n^{(1)}+\cdots$, $|\Psi_n\rangle=|\Psi^{(0)}_n\rangle+\lambda|\Psi^{(1)}_n\rangle+\cdots$, where $|\Psi^{(0)}_n\rangle$ and $E_n^{(0)}$ are the eigenvalues and eigenvectors of $H_0$. To attain
the degree of accuracy characterizing the classical resonant approximation, corrections to $|\Psi^{(0)}_n\rangle$ are, in fact, completely irrelevant. They never grow and never reach a magnitude of $O(1)$. Corrections to $E_n$ on the other hand, may be relevant, since they are multiplied by $t$ in (\ref{Psieig1}) and become of order 1 when $t\sim 1/\lambda$. However, by this reasoning, only the first correction $E_1$ matters (higher corrections may only enter the game on much longer time scales like $1/\lambda^2$). Thus, the classical resonant approximation corresponds to the quantum evolution
in which the energy eigenstates remain uncorrected, and the corresponding eigenvalues are corrected at order $\lambda$:
\beq
|\Psi(t)\rangle_{res}=\sum_n  e^{-i(E_n^{(0)}+\lambda E_n^{(1)}) t} \langle \Psi^{(0)}_n|\Psi(0)\rangle\,|\Psi^{(0)}_n\rangle.
\label{Psieig}
\eeq

The computation of the energy shifts $E_n^{(1)}$ is completely standard and treated in any textbook on quantum mechanics.
They are expressed through the matrix elements of the perturbation Hamiltonian $H_{int}$ in the eigenbasis of $H_0$.
One essential point is that for a field theory with a resonant spectrum of normal mode frequencies, as in (\ref{eq_eigenvalueAdS}),
the spectrum of energies of the free Hamiltonian is highly degenerate. One thus has to deal with the energy shifts in a degenerate spectrum,
again by completely standard textbook methods of quantum mechanics. The quantum version of the resonant Hamiltonian (\ref{Hresclas})
will automatically emerge from this analysis, confirming our expectation that the leading order energy shifts of the quantum theory
correctly capture the content of the classical resonant approximation.

\subsection{Quantum perturbation theory in AdS}

The quantum Hamiltonian corresponding to (\ref{Lagr}) is
\beq\label{H}
{\cal H} =  \int  \Big[\frac{\pi_\phi^\dagger\pi_\phi}{\cos^2\hspace{-0.7mm}x\hspace{1mm}} + \cos^2\hspace{-0.7mm}x\hspace{1mm} \nabla\phi^\dagger\cdot\nabla\phi+m^2\phi^\dagger\phi+\frac{\lambda\phi^{\dagger 2}\phi^2}{2}\Big] \,d^d {\bf x},
\eeq
where the Schr\"odinger picture operators only depend on the spatial coordinates denoted collectively as ${\bf x}=(x,\Omega)$, and $\pi_\phi$ is the momentum field conjugate to $\phi$. The volume element is $d^d {\bf x}=dx\,d\Omega \tan^{d-1}x/\cos^2x$ and the dot-product is computed with the $d$-sphere metric $dx^2+\sin^2 x \, d\Omega^2_{d-1}$. For each normal mode function of $\phi$ \eqref{eq_modefunctionAdS} we introduce a pair of creation-annihilation operators for the particles and antiparticles, and expand the field and conjugate momentum as follows
\beq
\phi=\sum_{nlk}\frac1{\sqrt{2\om_{nlk}}}\left(a_{nlk}+b^\dagger_{nlk}\right) e_{nlk}({\bf x}),\qquad {\pi_\phi}=i\sum_{nlk}\sqrt{\frac{\om_{nlk}}{2}}\left(a^\dagger_{nlk}-b_{nlk}\right) \cos^{2}\hspace{-0.7mm}x\hspace{1mm} \bar{e}_{nlk}({\bf x}),\label{phiexp}
\eeq
so that
\begin{align}
[a^\dagger_{nlk},a_{n'l'k'}]=-\de_{nn'}\de_{ll'}\de_{kk'}, \\
[b^\dagger_{nlk},b_{n'l'k'}]=-\de_{nn'}\de_{ll'}\de_{kk'}.
\end{align}
The mode functions are normalized as
\beq
(e_{nlk} e^{-i \om_{nlk}t},e_{n'l'k'} e^{-i \om_{n'l'k'}t}) = 2\om_{nlk}\de_{nn'}\de_{ll'}\de_{kk'}
\eeq
with respect to the Klein-Gordon inner product
\beq 
(\phi_1,\phi_2)=-i \int_{\Sigma} d\Sigma^\mu \sqrt{-g_{\Sigma}}\left( \phi_1(x)\partial_\mu \bar{\phi}_2(x) - (\partial_\mu\phi_1(x)) \bar{\phi}_2(x)\right),
\eeq
where $\Sigma$ is a constant time slice, $d\Sigma^\mu = n^\mu d\Sigma$ and $n_\mu =\frac1{|\cos x|} \partial_t$ is a future-directed unit vector normal to $\Sigma$.
The Hamiltonian (\ref{H}) can be expressed as a function of the creation-annihilation operators (subtracting the ground state energy)
\beq
{\cal H}= \sum_{nlk}\om_{nlk}( a^\dagger_{nlk}a_{nlk} +b^\dagger_{nlk}b_{nlk} )+ \lambda\,\hat{\cal H}_{int} ,\label{Hdecmp}
\eeq
where $\hat{\cal H}_{int}$ is simply $\frac12 \int \phi^2 \phi^{\dagger2}$ expressed through (\ref{phiexp}), 
\beq
{\cal H}_{int}= \frac12 \hspace{-1mm}\sum_{\eta_1\eta_2\eta_3\eta_4} \hspace{-1mm}C_{\eta_1\eta_2\eta_3\eta_4}:(a^\dagger_{\eta_1}+b_{\eta_1})(a^\dagger_{\eta_2}+b_{\eta_2})(a_{\eta_3}+b^\dagger_{\eta_3})(a_{\eta_4}+b^\dagger_{\eta_4}):\,.\label{Hintdef}
\eeq
The colons indicate normal ordering, and the interaction coefficients $C$ are given by (\ref{Cgen}). We have introduced the collective index $\eta=(n,l,k)$ so that $\sum_\eta=\sum_{n,l,k}$, $\om_\eta=\om_{nlk}$, $\eta_1=(n_1,l_1,k_1)$, etc. 

For $\lambda=0$, the system reduces to a collection of decoupled oscillators where the eigenstates are Fock states $|\{n,\tilde n\}\rangle$ with $n_\eta$ particles and $\tilde n_\eta$ antiparticles for each AdS normal mode $\eta$:
\beq
a^\dagger_\eta	a_\eta|\{n,\tilde n\}\rangle = n_\eta |\{n,\tilde n\}\rangle,\qquad b^\dagger_\eta b_\eta|\{n,\tilde n\}\rangle = \tilde n_\eta |\{n,\tilde n\}\rangle.
\label{occupnum}
\eeq
The energy is
\beq\label{Edef}
E=\sum_\eta \om_\eta (n_\eta+\tilde n_\eta).
\eeq
Note that the spectrum of the non-interacting theory is highly degenerate. Introducing the total particle and antiparticle numbers
\beq
N=\sum_\eta n_\eta,\qquad \tilde N=\sum_\eta \tilde n_\eta,
\eeq
we note that
\beq\label{Epdef}
E'=E-\de (N+\tilde N)
\eeq
is an integer, being made of integers $\om_\eta-\de$, $n_\eta$ and $\tilde n_\eta$. Of course, there are many different ways to generate a given integer value of $E'$ by choosing $n_\eta$ and $\tilde n_\eta$, which results in huge degeneracies (the level multiplicities grow without bound as one moves to higher energies). 

When a small nonzero $\lambda$ is turned on in (\ref{Hintdef}), the degenerate energy levels we have just described split. To analyze this splitting at linear order in $\lambda$ one must simply compute the matrix elements of $\hat{\cal H}_{int}$ between the states {\it within the same} unperturbed energy level, and diagonalize the resulting finite-sized matrices. Before proceeding with these matrix elements, we must characterize the degeneracies more explicitly. If $\de$ is a generic real number, different values of $N+\tilde N$ always correspond to different energies. Since the Hamiltonian exactly conserves $N-\tilde N$, this means that the matrix elements $\langle \Psi'|{\cal H}_{int}|\Psi\rangle$ are only nonzero if $|\Psi'\rangle$ has the same number of particles as $|\Psi\rangle$, and the same for antiparticles. This pattern, in fact, persists even if $\de$ takes special rational values (as it does for massless fields) so that there are degeneracies between levels with different $N+\tilde N$. The reason lies in the selection rules of \cite{Yang,EN}, which make the interaction coefficients $C$ vanish for any term in (\ref{Hintdef}) that could contribute to such matrix elements connecting different values of $N+\tilde N$. For example, there are terms in (\ref{Hintdef}) of the form $a^\dagger_{\eta_1} a^\dagger_{\eta_2}b^\dagger_{\eta_3}a_{\eta_4}$. Such term, if it contributed, would generate nonzero matrix elements of ${\cal H}_{int}$ between states with $N+\tilde N$ differing by 2. However, this term cannot contribute: because we are only computing matrix elements within the same unperturbed energy level, the energy $E$ given by (\ref{Edef}) must be the same for $|\Psi\rangle$ and $|\Psi'\rangle$ in $\langle \Psi'|{\cal H}_{int}|\Psi\rangle$. In order for this to be true, one must have $\om_4=\om_1+\om_2+\om_3$. But the interaction coefficients corresponding to mode quartets with such frequency relations precisely vanish due to the selection rules of \cite{Yang, EN}. Similar considerations hold for any other terms in (\ref{Hintdef}) that change the number of particles or antiparticles. One may summarize that {\it particle production plays no role in the computation of the energy splitting at leading order in} $\lambda$.

In view of the above structure of the matrix elements of (\ref{Hintdef}), the computation proceeds independently in each block indexed by three integers $E'$, $N$ and $\tilde N$. In fact, in correspondence with the consistent truncations we performed in classical theory, we prefer to focus here on blocks with
\beq
\tilde N=0,
\eeq
i.e., blocks without antiparticles. Then, for $|\Psi\rangle$ and $|\Psi'\rangle$ without antiparticles, one may write
\beq
\langle \Psi'|{\cal H}_{int}|\Psi\rangle=\frac12 \hspace{-3mm}\sum_{\eta_1,\eta_2,\eta_3,\eta_4} \hspace{-3mm}C_{\eta_1\eta_2\eta_3\eta_4}\langle \Psi'| a^\dagger_{\eta_1}a^\dagger_{\eta_2} a_{\eta_3} a_{\eta_4}|\Psi\rangle.
\eeq
But since we are only interested in $|\Psi\rangle$ and $|\Psi'\rangle$ with the same energy, only terms with $\om_1+\om_2=\om_3+\om_4$ may contribute, and hence
\beq
\langle \Psi'|{\cal H}_{int}|\Psi\rangle=\langle \Psi'|{\cal H}_{res}|\Psi\rangle,
\eeq
where
\beq
{\cal H}_{res}=\frac12\hspace{-1mm}\sum_{\om_1+\om_2=\om_3+\om_4} \hspace{-4mm}C_{\eta_1\eta_2\eta_3\eta_4}a^\dagger_{\eta_1}a^\dagger_{\eta_2}a_{\eta_3}a_{\eta_4}.
\label{quantresful}
\eeq
This is precisely the Hamiltonian corresponding to the resonant system (\ref{alrestrunc}). 

Note furthermore that, by angular momentum conservation, (\ref{quantresful}) can only have nonzero matrix elements between states that carry the same total amount of angular momentum projection on the polar axis. The Hamiltonian diagonalization and energy corrections can thus be discussed independently
for such sectors with different values of total angular momentum. In what follows, we shall focus on the sectors with the maximal value of the angular momentum projection on the polar axis for a given number of particles and a given unperturbed energy.


\subsection{Maximally rotating sector in the quantum theory} \label{sec:Maximally rotating sector in the quantum theory}

We have seen in section~\ref{secmxrota} that the classical resonant system (\ref{alrestrunc}), which is a counterpart of (\ref{quantresful}), can be consistently truncated to the set of modes that carry the maximal amount of angular momentum (in terms of its projection on the polar axis) for a given normal mode frequency. If the initial state of the classical evolution (\ref{alrestrunc}) only has such modes excited, no modes outside this sector will ever get excited at later times. What does this consistent truncation imply for the corresponding quantum theory (\ref{quantresful})?

Classical truncations to subsets of degrees of freedom in general do not have immediate implications for the corresponding quantum theory.
Indeed, classically, one may simply set to zero a selection of degrees of freedom (both coordinates and momenta) in the initial state and observe that they do not get excited by the subsequent evolution. The same cannot literally be done in the corresponding quantum theory, since the uncertainty principle
forbids setting both coordinates and momenta to zero. A straightforward example in our context is the spherically symmetric truncation, which keeps
only the AdS modes of zero angular momentum. This truncation is always classically consistent because of the rotational symmetry of the equations of motion, and has in fact been very commonly employed in studies of classical dynamics of AdS perturbations \cite{BR,AdSLMS,FPU,CEV,CEV2,BKS,revival,DKFK,DFLY,BMR,GMLL,BR2,returns,CF}. Still, the existence of these truncations does not translate directly to the corresponding quantum theory. Namely, if one wants to compute corrections to energies of multiparticle states with zero angular momentum, one must still deal with the full quantum resonant Hamiltonian (\ref{quantresful}), and cannot recover these corrections from a Hamitonian in which all non-spherically-symmetric modes have been simply discarded. Of course, as one approaches the semi-classical regime, the classical consistent truncation must still manifest itself and become valid approximately, but it is never valid exactly in the quantum theory.

The situation, however, is much simpler and more tractable for the maximally rotating truncations that we focus upon here.
The reason for this is largely kinematic and originates from the location of the maximally rotating states at the boundary of the tower
of modes in Fig.~\ref{figtower}, which makes it difficult for these states to interact with other states via interactions that conserve energy and angular momentum. This is somewhat reminiscent of the simplification of interactions near the Fermi surface in solid state physics.

To see how the decoupling of the maximally rotating sector happens in practice within the quantum theory, consider first the composition
of maximally rotating multiparticle states. A state with $N$ particles is maximally rotating if it has $M$ units of angular momentum in projection on the polar axis (we choose $M$ to be positive as a matter of convention) and $N\de+M$ units of energy.  Since the minimal value of energy of an individual particle with angular momentum $m$ is $\de+|m|$, and energies as well as angular momenta add when creating 
multiparticle states, it means that the only way a multiparticle state can be maximally rotating is that all the individual particles occupy maximally rotating states. (The multiplets where these states reside are highlighted in Fig.~\ref{figtower}, and there is only one such state for each frequency level.) In other words, the occupation numbers $n_\eta$ in (\ref{occupnum}) are only nonzero if $\eta$ is a maximally rotating mode with angular momentum projection $m>0$ and energy $\de+m$. (As a reminder, we only consider states whose `antiparticle' occupation numbers $\tilde n_\eta$ in (\ref{occupnum}) are zero.) Due to this structure of mode occupation, when computing the matrix elements of (\ref{quantresful}) between such maximally rotating multiparticle states, only terms contribute where $\eta_1$, $\eta_2$, $\eta_3$, $\eta_4$ all represent maximally rotating modes. Hence, one can simply discard all the non-maximally-rotating modes
for the purpose of the computation of energy shifts in the maximally rotating sector, which is precisely
what one does in the naive classical truncation. The energy shifts in the maximally rotating sector
can be computed exactly (at leading order in the coupling parameter) by considering the quantized
version of the classical maximally rotating Hamiltonian (\ref{Hresclas}), that is
\beq
H_{res}=\frac12 \sum_{n+m=k+l} C_{nmkl}\,a^\dagger_n a^\dagger_m a_k a_l,
\label{Hresq}
\eeq
with $C$ given by (\ref{Cmaxrt}), $[a^\dagger_m,a_n]=-\de_{mn}$, so that $a^\dagger_m$ and $a_m$ are the creation-annihilation operators for the unique
maximally rotating mode with $m$ units of angular momentum (in terms of its projection on the chosen rotation axis) and frequency $\de+m$.

The analysis of energy eigenstates of (\ref{Hresq}), which is the main technical objective of our treatment, follows the general pattern outlined in \cite{quantres}. The Hamiltonian (\ref{Hresq}) commutes with
\beq
N=\sum_{n=0}^{\infty} a^\dagger_n a_n,\qquad M=\sum_{n=1}^{\infty} n a^\dagger_n a_n,
\eeq
and the three operators can be diagonalized simultaneously. The eigenvalues of $N$ and $M$ are integers
that split the spectrum into $(N,M)$-blocks that can be treated independently. Furthermore, each such block is finite-dimensional, with dimensionalities related to integer partition numbers \cite{quantres}, since there is only a finite number of ways to satisfy $\sum_k n_k=N$ and $\sum_k k n_k=M$ with integer $n_k$. Diagonalization of (\ref{Hresq}) then reduces to diagonalizing a finite-sized numerical matrix at each value
of $N$ and $M$, and produces the eigenvalues $\varepsilon_I$ of (\ref{Hresq}), which provide the first order corrections to the degenerate unperturbed free energies of (\ref{H}) in the maximally rotating sector
\begin{equation}
    E_I^{(0)} + \lambda E_{I}^{(1)} = N\delta +M+\frac{\G{2\de-\frac{d}{2}}\G{\de}^2}{4 \pi^{d/2} \G{\de-\frac{d}{2}+1}^2 \G{2\de}}\lambda \varepsilon_I,
\end{equation}
where $I$ labels the different eigenstates of (\ref{Hresq}) within the same $(N,M)$-block.

There is a further conserved quantity
\beq
Z = \sum_{n=0}^\infty \sqrt{(n+1)(n+\delta)} a^\dagger_{n+1} a_n,
\eeq
inherited from (\ref{Zclas}). Note that $N$ manifestly commutes with all the conserved operators while $Z$ and $Z^\dagger$ respectively act as a raising and lowering operator for $M$
\beq
[M,Z] = Z,\qquad [M,Z^\dagger] = -Z^\dagger.
\label{eq: commutation relation Z,M}
\eeq
The commutation relation between $Z$ and $Z^\dagger$ involves both $M$ and $N$,
\begin{equation}
\left[Z,Z^\dagger \right] = - \delta N - 2M.
\label{eq: commutation relation Z}
\end{equation}
The operators $Z$ and $Z^\dagger$ play an essential role in the diagonalization procedure of \eqref{Hresq}. In particular, the entire spectrum can be easily reconstructed from states in the kernel of $Z^\dagger$ defined by
\beq
Z^\dagger|\Psi\rangle=0.
\label{Zker}
\eeq
(This condition physically implies that the center-of-mass of the corresponding multiparticle state is in its ground state.)
Indeed, since $Z$ and $Z^\dagger$ commute with the Hamiltonian \eqref{Hresq}, $Z$ and $Z^\dagger$ map  energy eigenstates in an ($N,M$)-block to energy eigenstates in the ($N,M+1$)-block and ($N,M-1$)-block, respectively, without changing the corresponding eigenvalue. Note that $Z$ can never annihilate a state since
\beq
\big|Z|\Psi\rangle\big|^2=\big|Z^\dagger|\Psi\rangle\big|^2+\langle\Psi|\de N +2M|\Psi\rangle\ge\langle\Psi|\de N+2M|\Psi\rangle,
\eeq
so that all eigenvalues in a given ($N,M$)-block are mapped to the next block.
In particular, this means that an arbitrary energy eigenvalue in some ($N,M$)-block must either originate from a lower $M$ block and then get transported by the action of $Z$ to the current block, or it must emerge as a new eigenvalue in the current block and be associated to an eigenvector $\ket{\psi}$ in the kernel of $Z^\dagger$. This latter statement follows directly from the fact that $Z^\dagger \ket{\psi}$ would otherwise be an eigenvector living in the ($N,M-1$)-block, which by assumption does not contain the associated eigenvalue in its spectrum.
In fact, it is straightforward to show that any vector $\ket{\phi}$ in the ($N,M$)-block that lies in the orthogonal complement of the $Z$-image of the ($N,M-1$)-block must be annihilated by $Z^\dagger$. One can simply write the condition $\braket{Z \psi |\phi} = 0$, which must hold for any state $\ket{\psi}$ in the ($N,M-1$)-block, as $\braket{ \psi | Z^\dagger | \phi} = 0$. This proves that $Z^\dagger \ket{\phi}$, a state in the ($N,M-1$)-block, is orthogonal to all states in that same block and hence must vanish. 

There is an efficient way to keep track of when a given eigenvalue first appeared in the spectrum by considering the product $Z Z^\dagger$. This operator is Hermitian and commutes with the Hamiltonian so that one can simultaneously diagonalize the two operators. As a consequence of the structure of eigenstates presented above, all the eigenstates of \eqref{Hresq} in the block ($N,M$) can be written as $Z^m \ket{\psi_{ker}}$ for some $m \geq 0$ and arbitrary $Z^\dagger$-kernel eigenstate $\ket{\psi_{ker}}$ in block ($N,M-m$). Using the commutation relations \eqref{eq: commutation relation Z,M} and \eqref{eq: commutation relation Z}, it is straightforward to show that such vectors are eigenstates of $Z Z^\dagger$ with eigenvalue 
\begin{equation}
Z Z^\dagger Z^m \ket{\psi_{ker}} = \left( m \de N+2 m(M-1)-m(m-1) \right)Z^m \ket{\psi_{ker}},
\label{eq: eigenvalues ZZdag}
\end{equation}
for $Z^m \ket{\psi_{ker}}$ in the ($N,M$)-block. All the vectors in the ($N,M$)-block that can be written in this way for a given value of $m$ therefore share the same eigenvalue under $Z Z^\dagger$. 

We therefore conclude that the simultaneous diagonalization of $Z Z^\dagger$ and $H_{res}$ in an ($N,M$)-block provides complete information about the spectrum in the current as well as in lower $M$-blocks and furthermore indicates the block in which the eigenvalues first appeared.


\section{Large $c$ holography, multiparticle states and their energy corrections}\label{sec:large}

Energy corrections for weakly nonlinear quantum fields in AdS fit very naturally in the framework of `large $c$ holography' \cite{largec1,largec2}, which points to their counterpart on the CFT side of holographic AdS/CFT dualities.

Isometries of the global AdS$_{d+1}$ space form the group $SO(d,2)$ which acts as conformal isometries on the conformal boundary of AdS given by the Einstein cylinder $R\times S^{d-1}$. The time translations
along the $R$-direction are the same as the AdS time translations, which gives a natural identification between the energies of AdS states and energies of CFT states on the boundary. If one then switches
to the Euclidean time evolution in the CFT and conformally maps $R\times S^{d-1}$ to $R^d$ in the manner of `radial quantization,' the time translations turn into dilatations in $R^d$ and the energies
of states are mapped into the conformal dimensions of the CFT operators, as per the usual operator-state correspondence. This construction gives an identification of energy eigenstates in the AdS bulk and operators of definite conformal dimension in the dual CFT.

For CFTs with holographic duals, the conformal dimensions must have a very special structure in order to match the patterns existing on the AdS side. Large $c$ holography is a concrete proposal as to how such correspondence can be implemented. One considers a family of CFTs of growing central charge $c$ and assumes that as $c$ goes to infinity, only a finite set of primary operators retain finite conformal dimensions. Furthermore, one assumes that these primaries attain the status of Generalized Free Fields (GFFs), which is expressed in simple factorization properties of their correlation functions, modelled on large $N$ factorization in gauge theories \cite{polyakov, makeenko}. For a generalized
free field $\cal O$, correlation functions satisfy  \cite{largec2}
\beq
\langle {\cal O}(x_1)\ldots {\cal O}(x_n)\rangle=\langle {\cal O}(x_1){\cal O}(x_2)\rangle \ldots \langle {\cal O}(x_{n-1}) {\cal O}(x_n)\rangle+\mbox{permutations},
\label{GFF}
\eeq
where one has to sum over all possible pairings. The fields thus behave as if they were free and their correlators satisfied Wick's theorem, but their conformal dimensions do not agree with the free value $(d-2)/2$. The above formula is understood to hold at $c=\infty$, and elsewhere receives corrections of order $1/c$. These corrections encode the interactions of the AdS side of the holographic duality.

The factorization property (\ref{GFF}) implies, in particular, that (at $c=\infty$) the conformal dimensions of GFFs and their descendants simply add up under multiplication \cite{largec2}. Hence, in this regime, one obtains an extremely simple spectrum of conformal dimensions for composite operators that matches, in fact,
the energy spectrum of free fields in AdS. Indeed, the descendants of a single GFF primary $\cal O$ of conformal dimension $\de$ are
of the form $\del_{m_1}\ldots \del_{m_p}{\cal O}$. Each differentiation increases the conformal dimension by 1,  giving a total dimension $\de+p$ and creating a tower of dimensions exactly identical to the frequency spectrum in AdS (which is the same as the single-particle energy spectrum) depicted in Fig.~\ref{figtower}. The level of the tower at $p$ units of energy above the ground state forms a rank $p$ fully symmetric tensor, and it can be decomposed into
angular momentum multiplets with $p=2n+l$ by separating this tensor into the traceless parts and the traces, so that individual dots in Fig.~\ref{figtower} correspond to states of the form $(\del^2)^n \del_{m_1}\cdots \del_{m_l}{\cal O}$, contracted with a fully symmetric fully traceless rank $l$ tensor whose indices are $m_1, ..., m_l$. (This matching between mode towers of free fields and descendants of a single CFT operator is treated at length in \cite{Terashima}.) Because the conformal dimensions of GFFs add under multiplication, as do energies of individual particles in multiparticle states of free fields, products of operators of the form $\del_{m_1}\cdots \del_{m_p}{\cal O}$ will possess a spectrum of conformal dimensions precisely matching the energies of
the Fock states\footnote{Many of such products are descendants rather than primary operators. The primaries, which cannot be expressed through derivatives of simpler operators, precisely correspond to combinations of Fock states in the kernel of the $Z^\dagger$ operator
of the bulk theory given by (\ref{Zker}).} of free particles in AdS, whose individual energies fit in the tower of Fig.~\ref{figtower}.
This scenario gives a concrete mechanism by which the AdS/CFT correspondence can work at $c=\infty$.
Its actual realization has been extensively discussed for higher spin holography in AdS$_3$ \cite{hspin,hspin2}.

As one moves away from the strict $c=\infty$ limit, the simple picture outlined above acquires $1/c$ corrections. In order for the match between the AdS and CFT sides of the holographic duality to persist
at order $1/c$, one must have the corrections to the conformal dimensions of multiparticle operators precisely reproduce the corrections to AdS energies described by the formalism of the previous section.
This, in particular, implies that the huge degeneracy of the spectrum at $c=\infty$ will be lifted by the $1/c$
corrections, in precisely the same manner as the multiparticle energy level degeneracy in AdS.

The language that we use here to discuss the AdS/CFT duality is slightly different from the one most commonly employed, namely, the language of the correlation functions. The correlation functions are expressed through the standard CFT data: the dimensions of the primary operators and their three-point correlation functions, which encode all the information about the CFT. We discuss instead the conformal dimensions of arbitrary multiparticle operators which contain full information about the Schr\"odinger evolution of arbitrary CFT states, and thus also provide a complete definition of the CFT. The two languages
must thus be equivalent. Discussions of conformal dimensions of operators have frequently appeared in
the context of AdS/CFT correspondence, occasionally adapted to the Hamiltonian perturbation theory for the
conformal dimensions, as in \cite{FSh}. These discussions most typically involve `small' operators made
from products of just a few primaries and their descendants, as in \cite{BSS,anomalous}, treated at the first
few orders of the perturbation theory at small coupling. By contrast, we restrict ourselves to the order $1/c$, but systematically deal with arbitrarily large operators. Indeed, one of our interests is establishing a connection to the classical dynamics in AdS, which, strictly speaking, involves taking the limit of an infinite
number of particles.

The main output of our considerations is an explicit family of energy eigenstates in AdS at order $\lambda$, whose individual particles reside in the maximally rotating sector of the spectrum marked in Fig.~\ref{figtower}. This translates in the holographically dual (large $c$) CFT to an explicit family of arbitrarily large operators
of definite conformal dimension (at order $1/c$) associated with the same part of the spectrum. The form
of these operators may be directly recovered from our explicit Fock space representation of the energy
eigenstates in AdS. The rest of our treatment aims at presenting, at a technical level, the construction of these states and their connection to the weakly nonlinear classical time periodicities in AdS.


\section{Energy ladders in the fine structure}\label{sec:rational}

We will now exploit the symmetries of the Hamiltonian \eqref{Hresq} and develop algebraic tools to derive the wavefunctions and associated energies of a specific family of eigenstates that we call {\it ladder states}. We will see that the occupation number operator $a^\dagger_n a_n$ evaluated in the ladder states at large $N$ and $M$ has features reminiscent of the classical ansatz \eqref{eq_adsansatz} for the modes $|\alpha_n|^2$. In the next section, we will make the connection to the classical analysis more concrete and build coherent-like combinations of the ladder states that reproduce the time-periodic features of the invariant manifold of classical solutions described in section~\ref{sectimeper}. The present analysis extends the techniques introduced in \cite{timeLLL}, applied to the simpler nonrelativistic version of the AdS system.

\begin{figure}[t]
\centering
\rule{0mm}{15mm}
\includegraphics[scale=1]{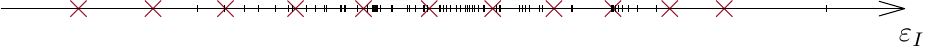}
\caption{The eigenvalues of \eqref{Hresq} in the block ($N=15,M=12$) at $\de=3$, computed by numerical diagonalization of the relevant Hamiltonian block. Generic eigenvalues are depicted as vertical bars and ladder state energies as red crosses. The lowest energy eigenvalue is a ladder eigenvalue, given by \eqref{eq: energy kernel ladder}, at $\varepsilon^{min}_{I}=57$ and the highest eigenvalue is given by $\varepsilon^{max}_{I}=N(N-1)/2=105$, which is the single eigenvalue originating from the block ($N,0$).\vspace{5mm}}
\label{fig: energies}
\end{figure}

We will start by showing that in every ($N,M$)-block with $ 2 \leq M \leq N$, there exists exactly one ladder state in the kernel of $Z^\dagger$, with energy
\begin{equation} \label{eq: energy kernel ladder}
\varepsilon^{(N,M)} = \frac{N(N-1)}{2}-\frac{N M \delta + M(M-1)}{2+4 \delta }.
\end{equation}
From the structure of the Hamiltonian blocks that we described in section \ref{sec:Maximally rotating sector in the quantum theory}, it follows that, in addition to the state with energy \eqref{eq: energy kernel ladder}, the block also contains $Z^m$-transported ladder states that were $Z^\dagger$-kernel eigenstates in the ($N,M-m$)-block with corresponding energy
\begin{equation}\label{eq: energy ladder}
\varepsilon_{(m)}^{N,M} =\frac{N(N-1)}{2} - \frac{N (M-m) \delta + (M-m)(M-m-1)}{2+4 \delta}.
\end{equation}
In summary, every ($N,M$)-block with $M \leq N$ contains $M-1$ ladder states, $M-2$ of which are transported from lower $M$ blocks. This structure of eigenvalues is illustrated in Fig.~\ref{fig: energies} for a block with $N=15$ and $M=12$. 

We now proceed with the construction of eigenstates with energy \eqref{eq: energy kernel ladder}. We start by considering the Hermitian operator 
\begin{equation}
B \equiv H_{res} - \frac{N(N-1)}{2} + \frac{1}{2+4 \delta }(N M \delta + M(M-1)-Z Z^\dagger),
\label{eq: definition B}
\end{equation}
which by definition annihilates a putative state with energy \eqref{eq: energy kernel ladder} that lies in the kernel of $Z^\dagger$. As a side note, it seems suggestive from the combination of \eqref{eq: classical eq for energy ladder1} and \eqref{eq: classical eq for energy ladder2} that an operator of this type would annihilate quantum states that are at the origin of the classical solutions. While the specific form of $B$ has been chosen by trial and error to enable our subsequent derivations, one guiding feature is that this operator remains purely quartic in the creation-annihilation operators when brought to the normal ordered form (and all the quadratic terms that could arise from the commutations cancel out).

Operator $B$ has a few further welcome features. First, it commutes with all the conserved operators, and specifically with $Z$ and $Z^\dagger$, as can be seen from
\begin{align}
[Z,Z Z^\dagger] = -Z \left( \delta N+2M \right), && [Z,\delta MN+M(M-1)] = Z \left( \delta N+2M \right).
\end{align}
Moreover, normal ordering the creation and annihilation operators reveals that it possesses the structure of a resonant Hamiltonian,
\begin{align}
B =  &H_{res}+ \frac{1}{2} \sum_{n,m=0}^\infty \left(\frac{n\de+nm}{1+2\de}-1\right)a^\dagger_n a^\dagger_m a_n a_m \nonumber \\
 &-\frac{1}{2+4\de} \sum_{n,m=0}^\infty \sqrt{(\de + m)(m+1)}\sqrt{(\de + n)(n+1)} a^\dagger_{n+1} a^\dagger_m a_n a_{m+1},
\end{align} 
which can be conveniently rewritten as a quadratic form
\begin{align}
&B= \frac{1}{2} \sum_{j=0}^{\infty} \sum_{k=0}^{j}A^\dagger_{jk} \Biggl[ \frac{\G{2\de}}{\G{\de}^2} \frac{\G{j+1}}{\G{j+2\de}} \sum_{l=0}^{j} A_{jl}+\frac1{2+4\de}\frac{k! (j-k)!}{\G{\de+k}\G{\de+j-k} }  \nonumber\\
&\times \Big( {\textstyle \left( j \de +2k(j-k)-\!2-4\de\! \right)\! A_{jk}-\!(k\!+\!1)(\!\de\!+\!j\!-\!k\!-\!1\!)A_{j,k+1}\!-\!(j\!-\!k\!+\!1)(\de+k-1) A_{j,k-1} }\Big)\Bigg] \nonumber \\
&\hspace{3mm}\equiv \sum_{j=0}^{\infty} \sum_{k,l=0}^{j} B^{(j)}_{kl} A^\dagger_{jk} A_{jl},
\label{eq: resonant definition B}
\end{align}
where we introduced $A_{jk} \equiv \sqrt{\frac{\G{\de+k}}{k!}}\sqrt{\frac{\G{\de+j-k}}{(j-k)!}} a_{k}a_{j-k}$ (with $A_{j,-1} \equiv 0 \equiv A_{j,j+1}$). 
We list a few relevant properties of the matrices $(B^{(j)})_{kl}$:
\begin{enumerate}
\item The matrices $(B^{(j)})_{kl}$ vanish for $j\le 3$.
\item At fixed $j$, we identify\footnote{One way to arrive at these null eigenvectors is to consider the classical counterpart of $B$ given by $H_{res}- \frac{N^2}2 + \frac{1+\de}{2+4\de} S^2$ with $S^2$ given by \eqref{eq: classical eq for energy ladder2}. 
This quantity can be represented in terms of quadratic forms defined by $B^{(j)}_{kl}$ in a way completely analogous to (\ref{eq: resonant definition B}). At the same time, substituting the values of $N$, $M$ and $Z$ for the ansatz (\ref{eq_adsansatz}) given by (\ref{eq: conserved quantities ansatz}-\ref{eq: conserved quantities ansatz 2}) into this classical expression for $B$ gives zero identically, as seen from combining \eqref{eq: classical eq for energy ladder1} and \eqref{eq: classical eq for energy ladder2},
which can be converted into explicit null eigenvectors for the individual $B^{(j)}_{kl}$, precisely of the form (\ref{zerov}).} two eigenvectors of the quadratic form $v^{(j)}_k B^{(j)}_{kl} v^{(j)}_{l}$ with zero eigenvalues
\begin{align}
&v^{(j)}_k=\frac{\G{\de+k}}{k!} \frac{\G{\de+j-k}}{(j-k)!} \label{zerov}\\ 
\mbox{and}\quad&v^{(j)}_k=\frac{\G{\de+k}}{(k-1)!} \frac{\G{\de+j-k}}{(j-k-1)!}.\nonumber
\end{align}
\end{enumerate}
Note that the operator $A_{jk}$ is invariant under $k\to j-k$. It is therefore sufficient to examine the properties of the quadratic forms at fixed $j$ \eqref{eq: resonant definition B} in the subspace where $v^{(j)}_k$ is identified with $v^{(j)}_{j-k}$. In these subspaces, we find:
\begin{enumerate}
  \setcounter{enumi}{2}
  \item Explicit diagonalization for multiple values of $j$ indicates that $B$ is a nonnegative operator with no other null directions than \eqref{zerov}. 
\end{enumerate}
As we already mentioned, $B$ annihilates a state in the kernel of $Z^\dagger$ with energy \eqref{eq: energy kernel ladder}. The construction of $Z^\dagger$-kernel ladder states therefore amounts to solving the following two equations simultaneously
\begin{align} \label{eq: kernel B}
B\ket{\psi} &= 0, \\
Z^\dagger\ket{\psi} &= 0,
\label{eq: kernel Z}
\end{align}
which, as we will now show, admits a single solution for every $ 2 \leq M \leq N$.  Note that property 3 listed above implies that the $Z^\dagger$-kernel ladder state is the lowest energy eigenstate in a given ($N,M$)-block. (Although we do not have an analytic proof of the nonnegativity of $B$, numerical diagonalization of the Hamiltonian strongly supports this statement.)

We start by making the crucial observation that property 1 implies that the states
\begin{align}
&\ket{N-M,M,0,0,0\dots}, \label{nilvec} \\
&\ket{N-M+1,M-2,1,0,0,0\dots}\nonumber
\end{align}
are annihilated by the operator $B$. These states therefore solve \eqref{eq: kernel B}; they do not solve \eqref{eq: kernel Z}, though. However, because $B$ commutes with $Z$ and $Z^\dagger$, one can generate a larger set of vectors satisfying \eqref{eq: kernel B} by applying $Z$ and $Z^\dagger$ repetitively,
\begin{equation}
\ket{\phi^{N,M}_m}=Z^m Z^{\dagger m}\ket{N-M,M,0,\dots},
\label{eq: building blocks ladder states}
\end{equation}
for $m=0 \dots M \leq N$, yielding a set of linearly independent vectors in the ($N,M$)-block. Their linear independence can be understood by noting that the operator $a^\dagger_{m+1} a_{m+1}$ has a nonzero expectation value in $\ket{\phi_m^{NM}}$ but annihlates all $\ket{\phi_n^{NM}}$ with $n<m$. Put differently, there is a nonzero term with mode number ${m+1}$ occupied in $\ket{\phi_m^{NM}}$, while this mode is absent in all states whose superposition makes up $\ket{\phi_n^{NM}}$ with $n<m$.
Note that the set of states \eqref{eq: building blocks ladder states} is equivalent to the set of states generated by $Z^m \ket{N-M+m,M-m,0,\dots}$ since 
\begin{equation}
Z^{\dagger m} \ket{N-M,M,0,\dots} = \sqrt{\frac{M!(N-M+m)!}{(N-M)!(M-m)!}} \sqrt{\de^m}  \ket{N-M+m,M-m,0,\dots}.
\label{eq: Zdagger action on O and 1}
\end{equation}

It turns out that these states are sufficient to construct a state that is annihilated by both $B$ and $Z^\dagger$. At fixed $M$ and $N$, we consider a linear combination of the vectors \eqref{eq: building blocks ladder states}
\begin{align}
\ket{\tilde{\psi}^{N}_M} = \sum^{M}_{m=0} b_m \ket{\phi^{NM}_m},
\label{eq: ladder state unknown bm}
\end{align}
and determine the coefficients $b_m$ by imposing that it should lie in the kernel of $Z^\dagger$. Using \eqref{eq: commutation relation Z} and \eqref{eq: Zdagger action on O and 1}, we note that
\begin{align}
Z^{\dagger} \ket{\phi^{N,M}_m} &= \left( Z^m Z^{\dagger m}   + m\left( \delta N + 2  M -(m+1)\right) Z^{m-1} Z^{\dagger m-1}  \right) Z^\dagger\ket{N-M,M,0,\dots} \nonumber \\
&= \sqrt{M(N-M+1)\delta} \left(\ket{\phi^{N,M-1}_m} + m\left( \delta N + 2  M -(m+1)\right) \ket{\phi^{N,M-1}_{m-1}}\right).
\end{align}
Requiring that $Z^\dagger$ should annihilate \eqref{eq: ladder state unknown bm} then provides the recursion relation
\begin{equation}
b_m = -\frac{b_{m-1}}{m \delta N + 2 m M-m(m+1)} ,
\end{equation}
which can be easily solved, setting $b_0 = 1$,
\begin{equation}
b_m = \frac{1}{m!(2 - 2 M - \delta N)_m}.
\end{equation}
We have therefore explicitly constructed an (unnormalized) eigenstate of \eqref{Hresq} 
\begin{equation}
\ket{\tilde{\psi}^{N}_M} =  \sum^{M}_{m=0} \frac{1}{m!(2 - 2 M - \delta N)_m} \ket{\phi^{N M}_m}
\label{eq: unnormalised kernel ladder state}
\end{equation}
in every ($N$,$M$)-block, where $2\leq M \leq N$. This state lies in the kernel of $Z^\dagger$, is annihilated by $B$ and has energy given by \eqref{eq: energy kernel ladder}, as can be simply read off from \eqref{eq: definition B}.

The norm of \eqref{eq: unnormalised kernel ladder state} can be computed by noting that most of the terms in the inner product are zero, because the $Z^\dagger$ operators contained in $\bra{\tilde{\psi}^{N}_M}$ annihilate $\ket{\tilde{\psi}^{N}_M}$,
\begin{align}
\braket{\tilde{\psi}^{N}_M|\tilde{\psi}^{N}_M} &=  \braket{N-M,M,0,\dots|\tilde{\psi}^{N}_M}= \sum_{m=0}^{M} \frac{\delta^m}{m!(2 - 2 M - \delta N)_m}  \frac{M!}{(M-m)!} \frac{(N-(M-m))!}{(N-M)!}\nonumber \\
&= \frac{M!}{(2 - 2 M - \delta N)_M} P_M^{( 
  1 - 2 M - \delta N, -1 + N + \delta N)}(1+2\delta),
  \label{eq: norm ladder state}
\end{align}
where $P_n^{(\alpha,\beta)}(x)$ is a Jacobi polynomial, and we made use of \eqref{eq: building blocks ladder states} and \eqref{eq: Zdagger action on O and 1}.
This defines the normalized ladder state 
  \begin{align}
\ket{\psi^{N}_M} \equiv \frac{\ket{\tilde{\psi}^{N}_M}}{\sqrt{\braket{\tilde{\psi}^{N}_M|\tilde{\psi}^{N}_M}}},
\label{laddernorm}
 \end{align}
  which lies in the kernel of $Z^\dagger$ at level $(N,M)$, for all $2\leq M \leq N$. In addition to the $Z^\dagger$-kernel ladder state, each block also contains $Z^m$-transported kernel ladder states from lower angular momentum blocks. Note that these states are also eigenstates of the Hamiltonian \eqref{Hresq} that are annihilated by $B$, because $Z$ and $B$ commute. Using the diagonalized expression of $Z Z^\dagger$ \eqref{eq: eigenvalues ZZdag}, it is straightforward to show that the energy of the (unnormalized) eigenstates
  \begin{equation}
 \ket{\psi^{(N,M)}_m} \equiv  Z^m \ket{\psi^N_{M-m}}
 \label{eq: general ladder state}
\end{equation} 
is given by \eqref{eq: energy ladder}. One can deduce the same by simply noting that their energy originates from a $Z^\dagger$-kernel energy eigenstate with energy \eqref{eq: energy kernel ladder} in block ($N,M-m$).


\subsection{Occupation numbers in ladder states}

We will show in the next section that the classical three-dimensional invariant manifold of solutions (\ref{eq_adsansatz}) emerges from particular coherent-like combinations of the ladder states in the quantum theory. In order to make contact with these classical solutions, it is practical to first compute the expectation value of the product $ a^\dagger_n a_n$ in kernel ladder states $\left\lbrace\ket{\psi^{N}_M}\right\rbrace$, which, as we will see, already contains interesting features that can be related to the classical ansatz \eqref{eq_adsansatz}. 

We start by writing out three useful identities for future reference. First, using the commutation relations \eqref{eq: commutation relation Z,M} and \eqref{eq: commutation relation Z}, one can show that powers of $Z$ and $Z^\dagger$ commute in the following way:
\begin{equation}
Z^{\dagger m'}Z^m = \sum_{k=0}^{min(m,m')} \frac{(-1)^k m! m'!}{k!(m-k)!(m'-k)!}(1-k+m-m'-2 M-\delta N)_k Z^{m-k}Z^{\dagger m'-k}.
\label{eq: commutation relation Z m' Zd m}
\end{equation}
Second, we will often use the following identity to commute an annihilation operator $a_n$ past several factors of $Z$,
\begin{equation}
a_n Z^m = \sum_{k=max(0,n-m)}^n Z^{m-n+k}a_k \sqrt{ \frac{n!}{k!}} \sqrt{\frac{\Gamma(\delta+n)}{\Gamma(\delta+k)}} \frac{m!}{(n-k)!(m-n+k)!}.
\label{eq: an through Zm}
\end{equation}
The last identity, valid on any state $\ket{\psi}$ that lies in the kernel of $Z^\dagger$, allows one to commute powers of $Z^\dagger$ past an annihilation operator
\begin{equation}
Z^{\dagger l} a_n \ket{\psi} = (-1)^l \sqrt{\frac{(n+l)!}{n!}} \sqrt{ \frac{\Gamma(\delta+n+l)}{\Gamma(\delta+n)}}a_{n+l} \ket{\psi}.
\label{eq: Zld through am}
\end{equation} 

The expectation value $\bra{\psi^{N}_{M}} a^\dagger_n a_n \ket{\psi^{N}_M}$ can be interpreted as computing the norm squared of the state $a_n \ket{\psi^{N}_M}$. We therefore first argue that the state $a_n \ket{\psi^{N}_M}$ can be expanded as a linear combination of ladder states in the ($N-1$, $M-n$)-block. 
In order to see this, we first drag $a_n$ through the $Z$ operators inside the kernel ladder state $\ket{\psi^{N}_M}$ using \eqref{eq: an through Zm}. 
Note that for each term inside $\ket{\psi^{N}_M}$, after the application of \eqref{eq: an through Zm} only two terms remain in the sum over $k$, containing either of the two annihilation operators $a_0$ and $a_1$. This follows from the fact that the operators act on a state which is only made out of the two corresponding creation operators, as can be seen from \eqref{eq: Zdagger action on O and 1}.
We are therefore able to write $a_n \ket{\psi^{N}_M}$ as a sum of states of the form $Z\ket{\phi}$ with $\ket{\phi}$ in the kernel of $B$. Because $Z$ and $B$ commute, $a_n \ket{\psi^{N}_M}$ also lies in the kernel of $B$ and is therefore a linear combination of ladder states living in the ($N-1$, $M-n$)-block, i.e.
\begin{equation}
a_n \ket{\psi^{N}_M} = \sum_{l=0}^{M-n} \frac{\bra{\psi^{N-1}_{M-n-l}} Z^{\dagger l} a_n\ket{\psi^{N}_M} }{\bra{\psi^{N-1}_{M-n-l}} Z^{\dagger l}Z^l \ket{\psi^{N-1}_{M-n-l}}}Z^l \ket{\psi^{N-1}_{M-n-l}}.
\end{equation}
The expectation value of $a^\dagger_n a_n$ in ladder states therefore reduces to
\beq\label{eq: formula amplitude spectrum} 
\bra{\psi^{N}_{M}}a^\dagger_n a_n\ket{\psi^{N}_M} = \sum^{M-n}_{l=0} \frac{\left|\bra{\psi^{N-1}_{M-n-l}}Z^{\dagger l} a_n\ket{\psi^{N}_M}\right|^2}{\bra{\psi^{N-1}_{M-n-l}} Z^{\dagger l}Z^l \ket{\psi^{N-1}_{M-n-l}}} 
=\sum^{M-n}_{l=0} \frac{\left|\bra{\tilde{\psi}^{N-1}_{M-n-l}}Z^{\dagger l} a_n\ket{\tilde{\psi}^{N}_M}\right|^2}{\braket{\tilde{\psi}^{N}_{M}|\tilde{\psi}^{N}_{M}} \bra{\tilde{\psi}^{N-1}_{M-n-l}} Z^{\dagger l} Z^{ l}\ket{\tilde{\psi}^{N-1}_{M-n-l}}}.
\eeq
The denominators in \eqref{eq: formula amplitude spectrum} are easily computed using \eqref{eq: commutation relation Z m' Zd m}, which leads to
\begin{equation}
\bra{\tilde{\psi}^{N-1}_{M-n-l}} Z^{\dagger l} Z^{ l}\ket{\tilde{\psi}^{N-1}_{M-n-l}}= \braket{\tilde{\psi}^{N-1}_{M-n-l}|\tilde{\psi}^{N-1}_{M-n-l}}(-1)^l l! \left(1-l-2(M-n)-\delta (N-1)\right)_l,
\label{eq: norm Z states}
\end{equation}
together with \eqref{eq: norm ladder state}.

We now turn to the numerators in \eqref{eq: formula amplitude spectrum} and compute $\bra{\tilde{\psi}^{N-1}_{M-n-l}}Z^{\dagger l} a_n\ket{\tilde{\psi}^{N}_M}$. We start by applying \eqref{eq: Zld through am}, from which follows that
\begin{equation}
\bra{\psi^{N-1}_{M-n-l}} Z^{\dagger l}a_n \ket{\psi^{N}_M} = (-1)^l \sqrt{\frac{(n+l)!(\delta+n)_l}{n!}} \bra{\psi^{N-1}_{M-n-l}} a_{n+l} \ket{\psi^{N}_M}.
\end{equation}
In the next section, we will be interested in studying combinations of these expectation values in the classical limit, so that it is useful to consider the limit of large $N,M$ with a fixed $\rho$ given by
\beq
\rho\equiv \frac{M}{N}.
\eeq
As we will see, a matrix element of the type $\bra{\psi^{N-1}_{M-n-l}} a_{n+l} \ket{\psi^{N}_M}$ scales with $\sqrt{N}$ in the classical limit, independently of the mode number. Therefore, terms with $l>0$ in \eqref{eq: formula amplitude spectrum} are suppressed
\begin{equation}
\frac{\left|\bra{\psi^{N-1}_{M-n-l}}Z^{\dagger l} a_n\ket{\psi^{N}_M}\right|^2}{\bra{\psi^{N-1}_{M-n-l}} Z^{\dagger l}Z^l \ket{\psi^{N-1}_{M-n-l}}} = \frac{(-1)^l (n+l)!(\delta+n)_l}{ n! l! \left(1-l-2(M-n)-\delta (N-1)\right)_l}\left|\bra{\psi^{N-1}_{M-n-l}} a_{n+l} \ket{\psi^{N}_M}\right|^2,
\end{equation}
and the $l=0$ term dominates the sum. In the classical limit, one therefore finds
\begin{equation}
\bra{\psi^{N}_{M}}a^\dagger_n a_n\ket{\psi^{N}_M} \approx \left|\bra{\psi^{N-1}_{M-n}} a_n\ket{\psi^{N}_M}\right|^2,
\end{equation}
as one could expect.

We proceed with the computation of $\bra{\psi^{N-1}_{M-n}} a_n\ket{\psi^{N}_M}$ and start by applying \eqref{eq: an through Zm} to commute $a_{n}$ past factors of $Z$ in the various terms in $\ket{\tilde{\psi}^{N}_M}$. Once again, only two terms remain in the sum over $k$ on the right-hand side of \eqref{eq: an through Zm}.
Moreover, because the bra state lies in the kernel of $Z^\dagger$, the only remaining terms in the expansion for the ket state \eqref{eq: ladder state unknown bm} that produce a nonzero contributions are those for which $m-n+k = 0$, with $k$ either 0 or 1. Taking this and \eqref{eq: norm Z states} into account, once the dust settles one obtains 
\begin{align}
\bra{\psi^{N-1}_{M-n}} a_n \ket{\psi^{N}_M} 
= &\frac{\sqrt{(\delta)_{n}\delta^{n}}}{\sqrt{n!}(2-2M-\delta N)_{n}} \left( N-M+n + \frac{n(1-2M-\delta N+n)}{\delta}\right) \nonumber \\ &\times   \sqrt{\frac{A(M-n,N-1) }{A(M,N)}},
\label{eq: exp value an in ladder state exact}
\end{align}
where $A(M,N)$ is defined as 
\begin{align}
A(M,N) &= \sum_{m=0}^{M} \frac{\delta^m}{m!(2 - 2M - \delta N)_m} \frac{(N-(M-m))!}{(M-m)!}\nonumber\\
&= \frac{(N-M)!}{(2-2M-\delta N)_{M}}P_{M}^{(1-2M-\delta N,-1+N(\delta+1))}(1+2\delta).
\end{align}

In the limit of large $N$, $M$ and fixed ratio $\rho= M/N$, the expectation value \eqref{eq: exp value an in ladder state exact} can be approximated as follows
\begin{align}
\bra{\psi^{N-1}_{M-n}} a_n\ket{\psi^{N}_M} \approx \frac{\sqrt{N}}{\sqrt{1-\rho}} \frac{\sqrt{(\delta )_n}}{\sqrt{n!}} \left(\frac{\delta (\rho-1)}{\delta+\rho}\right)^{n/2}\left(\frac{\delta+2\rho}{\delta+\rho}\right)^{\delta/2}\left( 1-\rho -\frac{n(2\rho+\delta)}{\delta}\right) \sqrt{R},
\label{eq: exp value an in ladder state s1}
\end{align}
where $R$ is the following ratio of Jacobi polynomials
\begin{align} \label{eq: ratio jacobi}
R = \frac{P_{M-n}^{(1-2(M-n)-\delta (N-1),-1+(N-1)(\delta+1))}(1+2\delta)}{P_M^{(1-2M-\delta N,-1+N(\delta+1))}(1+2\delta)}.
\end{align}
The limiting behavior of this ratio at large $N$ and $M = \rho N$ can be computed using \cite{jacobi} (see appendix \ref{app: Asymptotic behavior jacobis}) and yields
\begin{align}
R = \frac{(\rho-1) (-1)^{n}  \left( \gamma^{-1}
   \left(\frac{\delta}{\rho}+2\right)+\delta+2\right)^{-n}}{\left( \rho \gamma-1\right) \left((\de +1)^{-1}\gamma^{-1}+1\right)^{\delta}},
\end{align}
with the factor $\gamma$ rewritten as a function of $\rho$, 
\begin{equation}
\gamma = \sqrt{\frac{\rho+\de}{\rho\left(1+\de\right)}}.
\label{eq: definition gamma quantum}
\end{equation}
We insert this in \eqref{eq: exp value an in ladder state s1} and find that, in the classical limit, 
\begin{align}
\bra{\psi^{N-1}_{M-n}} a_n\ket{\psi^{N}_M} \approx & \frac{\sqrt{N}}{\sqrt{1-\rho \gamma}} \frac{\sqrt{(\delta)_n}}{\sqrt{n!}} \left(\frac{\delta (1-\rho)}{(\delta+\rho)(2+\delta+\gamma^{-1}\left(2+\frac{\delta}{\rho}\right))}\right)^{n/2} \nonumber \\
&\times \left(\frac{\delta +2\rho}{\rho \gamma+\rho+\delta}\right)^{\delta/2}\left( 1-\rho -\frac{n(2\rho+\delta)}{\delta}\right), \\ 
\equiv & f_n (b + an)p^n
\label{eq: exp value an in ladder state limit}
\end{align}
which demonstrates that the ladder state reproduces the form of the classical ansatz \eqref{eq_adsansatz}. Note however that the time evolution of ladder states is too simple to reproduce the time periodicities found in the classical analysis, as they are eigenstates of the Hamiltonian \eqref{Hresq} and hence do not directly encode the dynamics of classical solutions. 


\section{Coherent states and classical time-periodic dynamics}\label{sec:coherent}

The properties of the ladder states derived in the previous section are reminiscent of the classical time-periodic solutions discussed in section \ref{sectimeper}. We will now complete the connection between the classical and the quantum theory by defining explicit coherent-like combinations of the ladder states that reproduce the periodic returns of the configuration of classical amplitudes described in section \ref{sectimeper}. 

Our construction of coherent-like states is inspired by the standard coherent states $\ket{\alpha}$ for the harmonic oscillator, which are a normalized infinite sum of terms $a^{\dagger N} \ket{0}$ with weight $\alpha^N e^{-|\alpha|^2/2}/\sqrt{N!}$ such that $a \ket{\alpha} = \alpha \ket{\alpha}$. Although a coherent state in principle contains an infinite number of terms with different energies and number of particles, at large values of $|\alpha|$, it is able to reproduce classical physics because only terms for which $N \sim |\alpha|^2$ are able to compete with the exponential suppression in $|\alpha|^2$ so that expectation values are dominated by a subset of high energy terms with energy $\sim \omega N \sim \omega |\alpha|^2$ and particle number $|\alpha|^2$. In the present case, we aim at reproducing a three-complex-parameter family of nonlinear classical periodic solutions using ladder states. This set of energy eigenstates similarly has an unbounded parameter $N$ and we will introduce a first complex parameter $\alpha$ which, when large, will tune which set of ladder states dominate the $N$-summation. In contrast to the harmonic oscillator modes, the kernel ladder states have an additional parameter $M$ which is bounded by $N$. We will therefore introduce a second weighted sum, running over $M$, and choose a distribution that peaks around a set of terms at large $N$ so as to reproduce the angular momentum of classical solutions. A natural choice for this is the binomial distribution which introduces one additional complex parameter $\beta$. The building blocks of our construction are the $Z^\dagger$-kernel ladder states \eqref{laddernorm}, so that we naturally start by proposing a family of coherent states that have zero expectation value for the operator $Z$. In section~\ref{subsec:coherent}, we will describe in detail how the coherent states reproduce the subspace of classical solutions for which $Z=0$, as given by the amplitudes \crefrange{eq: y on Z is 0 manifold}{eq: b2 on Z0} and periods \eqref{eq: period returns} and \eqref{eq: period a b on Z0}. In section~\ref{subsec:Z}, we will move outside of the $Z=0$ manifold and analyze the $Z$-translated coherent states, which will introduce a third complex parameter $q$ in agreement with the three-dimensional classical invariant manifold of solutions. 


\subsection{Coherent states with $Z=0$}\label{subsec:coherent}

Consider the following combination of kernel ladder states, parametrized by two complex numbers $\alpha$ and $\beta$
\begin{equation}
\ket{\alpha,\beta} =  e^{- | \alpha |^2/2}\sum_{N=0}^\infty (1+|\beta|^2)^{-N/2} \frac{\alpha^N}{\sqrt{N!}} \sum_{M=2}^N \beta^M \sqrt{ \frac{N!}{M! (N-M)!}} \ket{\psi^{N}_M}.
\label{eq: definition coherent state}
\end{equation}
In this expression, $\alpha$ plays a similar role to the parameter that is typically tuning how many particles will dominate the coherent state, while the summation over $M$ contains a factor reminiscent of the binomial coefficient that is very sharply peaked at large $N$ (hence, at large $|\alpha|^2$). Consequently, only a subset of the terms in the second sum will contribute substantially to expectation values. Moreover, we will analyze the coherent state \eqref{eq: definition coherent state} in the regime where it becomes semiclassical, which, as we will demonstrate later on, corresponds to the regime where $|\alpha|$ is large and $| \beta |$ is of order one.

In order to connect the dynamics of \eqref{eq: definition coherent state} to the classical analysis of section \ref{sectimeper}, we evaluate the expectation value of $a_n$ in these coherent-like states,
\begin{align}\label{eq: an coherent state s1}
\bra{\alpha,\beta} a_n \ket{\alpha,\beta} = &e^{- | \alpha |^2} \sum_{N=1}^\infty \frac{| \alpha |^{2 N} \sqrt{N}/ \bar{\alpha}}{N!} (1+|\beta|^2)^{-N+1/2}  \\ \times &\sum_{M=n}^N \frac{|\beta|^{2 M}}{\bar{\beta}^{n}} \sqrt{\frac{N!}{M! (N-M)!}\frac{(N-1)!}{(M-n)! (N-1-(M-n))!}} 
\bra{\psi^{N-1}_{M-n}} a_n \ket{\psi^{N}_M}, \nonumber
\end{align}
where it has been taken into account that the expectation value of $a_n$ is only nonzero between states in blocks ($N-1,M-n$) and ($N,M$), as determined in \eqref{eq: exp value an in ladder state limit}.
For $|\alpha| \gg 1$, the sum over $N$ is dominated by terms with $N \sim |\alpha|^2$. In the second sum, terms centered around $M \sim N|\beta|^2/(1+|\beta|^2)$ dominate due to the binomial distribution. We therefore conclude that in the classical regime, the dominant terms have both $M$ and $N$ large, with ratio $M/N \sim |\beta|^2/(1+|\beta|^2)$, so that the approximation \eqref{eq: exp value an in ladder state limit} is valid. Moreover, in this limit, one can approximate
\begin{align}
\sqrt{\frac{N!}{M! (N-M)!}\frac{(N-1)!}{(M-n)! (N-1-(M-n))!}} \approx \frac{N!}{M! (N-M)!} \sqrt{1-\frac{M}{N}}\left(\frac{\frac{M}{N}}{1-\frac{M}{N}}\right)^{n/2},
\label{eq: recover binomial}
\end{align}
and the second sum in \eqref{eq: an coherent state s1} reduces to $\sum_M |\beta|^{2M} {N \choose M}\,\, f(M/N)$ for some function $f$. The binomial distribution has a sharp peak of width $2\sqrt{N}$ around the value $N |\beta|^2/(1+|\beta|^2)$, so that the deviations from this mean value are effectively restricted to lie within one standard deviation. As a consequence, the function $f(M/N)$ is roughly constant when evaluated at the leading terms  and can be approximated by $f\left(M/N\right) =  f(\left(|\beta|^2/(1+|\beta|^2)\right)$. It can therefore be taken outside of the sum over $M$, which is then easily evaluated and yields
\begin{align}
\bra{\alpha,\beta} a_n \ket{\alpha,\beta} &\approx  e^{- | \alpha |^2} \sum_{N=1}^\infty \frac{| \alpha |^{2 N} N / \bar{\alpha}}{N!} (1+|\beta|^2)^{1/2} \frac{1}{\bar{\beta}^{n}}  f\left(\frac{|\beta|^2}{1+|\beta|^2}\right).
    \label{eq: an coherent state s2}
\end{align}
This last sum can be performed exactly using
\begin{equation}
\sum_{N=1}^\infty \frac{N\xi^N }{N!}=\xi e^\xi.
\label{a_sum}
\end{equation}
Note that the factors in \eqref{eq: recover binomial} combine with a few remaining factors in \eqref{eq: an coherent state s2} in such a way that the expectation value of $a_n$ in the coherent states is actually very similar to the expectation value in ladder states in the limit of large $N$ and $M$. More precisely, one can write,
\begin{align}
\bra{\alpha,\beta} a_n \ket{\alpha,\beta} &\approx \sqrt{\frac{(\delta)_n}{n!}} \left( a n +b \right) p^n,
\label{eq: an in coherent state}
\end{align}
with
\begin{align}
a &= - \frac{\alpha}{\sqrt{1-\rho \gamma}} \left(\frac{\delta +2\rho}{\rho \gamma+\rho+\delta}\right)^{\delta/2} \frac{\de + 2\rho}{\delta}, \label{eq: a}
\\
b &=  \frac{\alpha}{\sqrt{1-\rho \gamma}} \left(\frac{\delta +2\rho}{\rho \gamma+\rho+\delta}\right)^{\delta/2} (1-\rho), \label{eq: b}
\\
p &= \frac{|\beta|}{\bar{\beta}} \left(\frac{\delta (1-\rho)}{(\delta+\rho)(2+\delta+\gamma^{-1}\left(2+\rho^{-1}\delta\right))}\right)^{1/2},
\label{eq: p}
\end{align}
with $\gamma$ defined as in \eqref{eq: definition gamma quantum} and
\begin{equation}
\rho = \frac{|\beta|^2}{1+|\beta|^2}.
\end{equation}
A similar computation can be worked out for $\bra{\alpha,\beta} a^\dagger_n a_n \ket{\alpha,\beta}$, where one finds
\begin{align}
\bra{\alpha,\beta} a^\dagger_na_n \ket{\alpha,\beta} &\approx \left| \bra{\alpha,\beta} a_n \ket{\alpha,\beta}\right|^2,
\end{align}
which can be used to show that 
\begin{align}
\bra{\alpha,\beta} N  \ket{\alpha,\beta} &= \left| \alpha \right|^2 \label{eq: N check}\\
\bra{\alpha,\beta} M  \ket{\alpha,\beta} &= \left| \alpha \right|^2 \frac{\left| \beta \right|^2}{1+\left| \beta \right|^2} \label{eq: M check}
\\ 
\bra{\alpha,\beta} Z \ket{\alpha,\beta} &= 0.
\label{eq: Z check}
\end{align}
Note that these equations can also be obtained using the classical formulas \eqref{eq: conserved quantities ansatz} and \eqref{eq: conserved quantities ansatz 2} expressing $N$, $M$ and $Z$ as a function of $a$, $b$ and $p$ and \crefrange{eq: a}{eq: p}. Hence, the absolute values of the complex parameters $\alpha$ and $\beta$ can be interpreted as tuning the particle number $N$ and the angular momentum (or energy) $M$ respectively. Because the coherent-like states \eqref{eq: definition coherent state} are made of kernel ladder states, evaluating $Z$ in those states evidently yields zero. In order to recover the full three-dimensional classical manifold of solutions, one must therefore consider the $Z$-translated state, $e^{\bar{q}Z-qZ^\dagger} \ket{\alpha,\beta}$, parametrized by a third complex number $q$. We postpone this discussion to the next subsection.

At this point, one can compare the classical (constant) amplitudes \crefrange{eq: p2 on Z is 0 manifold}{eq: b2 on Z0} on the manifold $Z=0$ with the parameters \crefrange{eq: a}{eq: p} which originate from expectation values in the coherent states and verify that these expressions agree upon the identifications \eqref{eq: N check} and \eqref{eq: M check}. This result is a first non-trivial verification that the states we constructed reproduce the classical solutions. We furthermore need to explore the time-dependence of the coherent states in order to compare it to the classical periods of $a$ and $b$ \eqref{eq: period a b on Z0} and the period of the function $p$ \eqref{eq: period returns} when $Z=0$, which are in turn related to the periodic returns \eqref{eq: period returns} of a general solution on the three-dimensional invariant manifold. 

We therefore turn to the time-evolution of the expectation value of $a_n$ in the coherent states \eqref{eq: an in coherent state} as induced by the resonant Hamiltonian \eqref{Hresq}:
\begin{equation}
\braket{ a_n }_\tau \equiv \bra{\alpha,\beta} e^{i H_{res}\tau} a_n  e^{-i H_{res}\tau} \ket{\alpha, \beta},
\end{equation} 
where we re-introduce the slow time $\tau = \lambda t$.
First, we recall that the non-vanishing terms in the expectation value of $a_n$ in a coherent state are of the type $\bra{\psi^{N-1}_{M-n}} a_n \ket{\psi^{N}_M}$. 
Those terms involve energy eigenstates with energy \eqref{eq: energy kernel ladder} and therefore evolve with a frequency that is linear in $N$ and $M$
\beq
\omega_{MN}=\varepsilon^{(N-1,M-n)}-\varepsilon^{(N,M)}=\frac{2+\delta(4-n)-n-n^2}{4\delta+2}+\frac{\delta +2n}{4\delta+2}M + \frac{\delta(n-4)-2}{4\delta+2} N.
\eeq
The time-evolution of \eqref{eq: an in coherent state} can then be computed by including the phase factors $e^{i\omega_{MN}\tau}$ in \eqref{eq: an coherent state s1} and working again through the summation over $M$, where the mean value of the distribution $\frac{|\beta|^2} {1+|\beta|^2}$ undergoes the following modification 
\begin{equation}
|\beta|^2 \rightarrow |\beta|^2 e^{\frac{\delta+2n}{4\delta+2} i \tau}. 
\label{eq: modification beta time ev}
\end{equation}
Before evaluating the $N$-summation, we pause and consider the classical limit in more detail by restoring the factors of $\hbar$. In the classical limit, measurable quantities such as the canonical field $\phi \sim \sum_n \sqrt{\frac{\hbar}{\omega_n}} a_n $ and its conjugate momentum field $\pi_\phi \sim \sum_n \sqrt{\hbar \omega_n} a_n $ should remain finite. One should therefore take $\hbar \rightarrow 0$ while keeping the product $|\langle a_n \rangle |^2 \hbar$ fixed. By requiring that the Hamiltonian \eqref{Hdecmp} should also remain finite as a function of the canonical variables, this effectively amounts to adding a factor of $\hbar^2$ in front of the Hamiltonian. Finally, by taking into consideration the usual factor of $\hbar$ in the denominator of the phase in the Schr\"odinger time evolution \eqref{Psieig}, one can account for the various factors of $\hbar$ by simply replacing $\tau$ by $\hbar \tau$ in all expressions. The modification \eqref{eq: modification beta time ev} to the mean value $\frac{|\beta|^2} {1+|\beta|^2}$ arising from the evaluation of the $M$-summation in the time-dependent case therefore becomes $|\beta|^2 \rightarrow |\beta|^2 e^{\frac{\delta+2n}{4\delta+2} i  \hbar \tau}$, which will have almost no noticeable effect in the classical limit for $\tau \sim 1$ (the exponential can in general be approximated to 1), except in the following factor 
\beq
 \left(\frac{1 +  |\beta|^2  e^{\frac{\delta+2n}{4\delta+2} i \hbar \tau}}{1 +  |\beta|^2}\right)^N \sim  \left(1+\frac{1}{N} \frac{|\beta|^2}{1+|\beta|^2}\frac{\delta+2n}{4\delta+2} i\hbar \tau N\right)^N\sim e^{i \hbar \tau \frac{\delta+2n}{4\delta+2} \frac{|\beta|^2} {1+|\beta|^2} N},
\eeq
where we used $\tau \sim 1$ in the first step while $N \hbar $ was kept fixed as $N \rightarrow \infty$ and $\hbar \rightarrow 0$ in the second step. 
In all the other places where the mean of the binomial distribution needs to be evaluated, one can safely ignore the time-correction to the mean value of the binomial distribution and simply evaluate it at $\frac{|\beta|^2} {1+|\beta|^2}N$ (because of the absence of any additional factor of $N$ that could compete with the presence of $\hbar$-factors). 
As a result, the $N$-summation can be performed using \eqref{a_sum} with $\xi = \left| \alpha \right|^2 e^{i \hbar \tau  \frac{(1+\left| \beta \right| ^2)(-2-4\delta+\delta n)+\left| \beta \right| ^2(\delta+2n)}{(4\delta+2)(1+\left| \beta \right| ^2)}} $ which results in the following time-dependence
\begin{align}
\bra{\alpha,\beta} a_n(t) \ket{\alpha,\beta} &\approx  \bra{\alpha,\beta} a_n(0) \ket{\alpha,\beta} e^{i\hbar \tau \frac{-n(1+n)+\left| \beta \right| ^2(\delta+n-n^2)}{(4\delta+2)(1+\left| \beta \right| ^2)}} e^{\left| \alpha \right| ^2\big( e^{i\hbar \tau  \frac{(1+\left| \beta \right| ^2)(-2-4\delta+\delta n)+\left| \beta \right| ^2(\delta+2n)}{(4\delta+2)(1+\left| \beta \right| ^2)}} -1\big) },\nonumber \\
&\approx  \bra{\alpha,\beta} a_n(0) \ket{\alpha,\beta} e^{\left| \alpha \right| ^2 i\hbar \tau  \frac{(1+\left| \beta \right| ^2)(-2-4\delta+\delta n)+\left| \beta \right| ^2(\delta+2n)}{(4\delta+2)(1+\left| \beta \right| ^2)} }. \label{eq: full time evolution an}
\end{align}
(Note that the first exponential in the first line can be simply approximated by 1 when $\hbar\tau$ is small.) Then at small $\hbar$ and finite $|\alpha|^2\hbar \tau$, separating the parts of the last exponent independent of $n$ and linear in $n$, we see that $a$ and $b$ share the same period $T_{ab}$ while $p$ evolves periodically with period $T$, where
\begin{align}
T_{ab} &= \frac{2 \pi (4\delta+2)(1+\left| \beta \right| ^2)}{\left| \alpha \right| ^2 \left((1+\left| \beta \right| ^2) (2+4\delta)-\left| \beta \right| ^2 \delta \right)}, \label{cohTab}\\
T &= \frac{2 \pi (4\delta+2)(1+\left| \beta \right| ^2)}{\left| \alpha \right| ^2 \left(\delta+ \left| \beta \right| ^2\delta +2\left| \beta \right| ^2 \right)}.
\label{eq: quantum period omega}
\end{align} 
Using \eqref{eq: N check} and \eqref{eq: M check}, these periods coincide with the classical periods \eqref{eq: period a b on Z0} and \eqref{eq: period returns} on the $Z=0$ slice of the three-dimensional invariant manifold. In the next subsection, we will discuss how the period of the parameter $p$ drives the non-trivial periodic returns for $Z\neq 0$ coherent states.

We conclude the discussion of the kernel coherent-like states by showing that the normalized standard deviation of the operator $a_n$ in the constructed states vanishes in the classical limit. Using \eqref{eq: full time evolution an},
\begin{align}
\braket{ a^\dagger_n a_n }_\tau - \left|\braket{  a_n }_\tau \right|^2 &=  | \braket{a_n}_0 | ^2 \left(1-e^{2| \alpha |^{2}\left( \cos\left(\left(\frac{(1+\left| \beta \right| ^2)(-2-4\delta+\delta n)+\left| \beta \right| ^2(\delta+2n)}{(4\delta+2)(1+\left| \beta \right| ^2)}\right)\hbar \tau\right)-1 \right)}\right), \nonumber\\
   &\approx | \braket{a_n}_0 | ^2 | \alpha |^{2} \left(\frac{(1+\left| \beta \right| ^2)(-2-4\delta+\delta n)+\left| \beta \right| ^2(\delta+2n)}{(4\delta+2)(1+\left| \beta \right| ^2)}\right)^2 \hbar^2 \tau^2.
\end{align}
We therefore find that, when the slow time $\tau$ is of order one, the standard deviation of $a_n$ is suppressed by a factor of $\hbar$ in the classical limit where $|\alpha|^2\hbar$ is kept finite,
\begin{align}
\frac{\braket{ a^\dagger_n a_n }_\tau - \left|\braket{  a_n }_\tau \right|^2}{\left|\braket{  a_n }_\tau \right|^2}   \sim \hbar\tau^2\cdot| \alpha |^{2} \hbar .
\end{align}
The states thus qualify as semiclassical.


\subsection{Z-translated coherent states}\label{subsec:Z}

The coherent-like states $\ket{\alpha,\beta}$ reproduce the dynamics of those classical solutions on the three-dimensional manifold (\ref{eq_adsansatz}) for which $Z$ vanishes identically. In order to reach the other solutions in this manifold, and recover the non-trivial periodic returns present in the amplitude spectrum of these solutions, one needs to map the coherent states \eqref{eq: definition coherent state} to states with nonvanishing $Z$. To this end, we define finite unitary symmetry transformations generated by $Z$ and act with them on $\ket{\alpha,\beta}$ as follows:
\begin{equation}
\ket{\alpha,\beta,q} = e^{\bar{q}Z-qZ^\dagger} \ket{\alpha,\beta}.
\label{eq: Ztranslated coherent states}
\end{equation}

Consider first
the infinitesimal versions of this transformation, converting a vector $\ket{\Psi}$ to $\ket{\tilde\Psi}=(1+\Delta\bar q Z- \Delta qZ^\dagger)\ket{\Psi}$, where $\Delta q$ is a small parameter.
From the commutation of $Z$ and $Z^\dagger$ with $a_n$, one easily recovers the expectation values of $a_n$ in $\ket{\tilde\Psi}$ as
\beq
\bra{\tilde\Psi} a_ n \ket{\tilde\Psi} = \bra{\Psi} a_ n \ket{\Psi} - \Delta q \sqrt{(n+1)(\delta+n)} \braket{\Psi|a_{n+1}|\Psi} + \Delta \bar{q} \sqrt{n(\delta+n-1)} \braket{\Psi|a_{n-1}|\Psi}.
\label{infZtr}
\eeq
As a consequence of $Z$ being bilinear in $a_n$ and $a^\dagger_n$, these transformations exactly
coincide with the corresponding classical transformations of $\alpha_n$ and $\bar\alpha_n$ previously treated 
in \cite{solvable}. Importantly, if $\bra{\Psi} a_ n \ket{\Psi}$ is in the form of the ansatz (\ref{eq_adsansatz}), one can straightforwardly verify that $\bra{\tilde\Psi} a_ n \ket{\tilde\Psi}$
respects that form as well, with infinitesimally shifted $a$, $b$ and $p$.

If one then starts applying such infinitesimal transformation consecutively to $\ket{\Psi}=\ket{\alpha,\be}$,
which respects the ansatz (\ref{eq_adsansatz}) at large $\alpha$ as seen from \crefrange{eq: an in coherent state}{eq: p}, it follows that the expectation values of $a_n$ in the transformed states will always respect the ansatz (\ref{eq_adsansatz}), which will also be true of $\braket{\al,\be,q|a_n|\al,\be,q}$ after the infinitesimal transformations have accumulated to the finite transformation \eqref{eq: Ztranslated coherent states}. Furthermore, since the infinitesimal transformations (\ref{infZtr}) exactly coincide with their classical analog, they can be explicitly integrated to finite transformations within the ansatz \eqref{eq_adsansatz},
as done in \cite{solvable} (the formulas below correct a small typo in \cite{solvable}). For imaginary $q = i \eta$, one gets
\begin{align}
& p\mapsto \frac{p - i\tanh\eta}{1 +i p \tanh\eta}, \label{transZ1}\\
& a\mapsto \frac{ap}{(p\cosh\eta - i \sinh\eta)(\cosh\eta + ip \sinh\eta)^{\de+1}},\label{transZ2}\\
& b \mapsto  \frac{ b(1 +i p \tanh\eta) -i  a \de p \tanh\eta}{(1 +i p \tanh\eta)(\cosh\eta + i p \sinh\eta)^\de} ,\label{transZ3}
\end{align}
and for real $q = \xi$,
\begin{align}
& p\mapsto \frac{p+\tanh{\xi}}{1+p \tanh{\xi} },\label{transZ4}\\
& a\mapsto \frac{ap}{(p\cosh\xi + \sinh\xi)(\cosh{\xi}+p \sinh{\xi} )^{\de+1}},\label{transZ5}\\
& b \mapsto \frac{b(1+p \tanh{\xi})-a\de p \tanh{\xi}}{(1+p \tanh{\xi})(\cosh{\xi}+p \sinh{\xi})^\de}\label{transZ6} .
\end{align}
By substituting the values of $a$, $b$, $p$ of a large $\alpha$ kernel coherent state $\ket{\al,\be}$,
given by \crefrange{eq: a}{eq: p}, one obtains the corresponding values for the $Z$-transformed coherent
state (\ref{eq: Ztranslated coherent states}). The expectation value of $a_n$ in (\ref{eq: Ztranslated coherent states}) can then be read off the ansatz (\ref{eq_adsansatz}).

Since $Z$ and $H$ commute, to compute the time evolution of $a_n$ in the coherent states $\ket{\al,\be,q}$ it suffices to apply the $Z$-transformation with a given $q$ to the time evolution of $a_n$ given by
(\ref{eq: full time evolution an}) and characterized by the periods \crefrange{cohTab}{eq: quantum period omega}. In view of \eqref{eq: an in coherent state}, the expectation value (\ref{eq: full time evolution an})
always fits in the ansatz (\ref{eq_adsansatz}), and can therefore be $Z$-transformed by applying \crefrange{transZ1}{transZ3} or \crefrange{transZ4}{transZ6}. In the state $\ket{\al,\be}$, $a$ and $b$ evolve by pure phase rotation with period \eqref{cohTab} and $p$ evolves by pure phase rotation with period \eqref{eq: quantum period omega}. By \crefrange{transZ1}{transZ6}, this implies that the absolute values of $a$ and $b$ are no longer constant in $\ket{\al,\be,q}$, but
$a$, $b$, $\Re{(a\bar{b})}$ and therefore the amplitude spectrum $|\braket{\al,\be,q|a_n(t)|\al,\be,q}|^2$ of a general coherent state are exactly periodic with period controlled by the periodicity of $p$ within the corresponding kernel coherent state $\ket{\al,\be}$ given by \eqref{eq: quantum period omega}. If expressed through the conserved quantities \crefrange{eq: N check}{eq: M check} in the state $\ket{\al,\be}$, which we denote $N$ and $M_0$, this period becomes
\beq
T=\frac{4\pi(1+2\de)}{N\de+2M_0}.
\label{TNM0}
\eeq
We still need to express this period through the expectation value of $M$ in the state $\ket{\al,\be,q}$,
rather than the state $\ket{\al,\be}$ from which $\ket{\al,\be,q}$ is obtained by \eqref{eq: Ztranslated coherent states}. The easiest way to do it is to notice that $S$ given by \eqref{eq: classical eq for energy ladder2} becomes $\sqrt{(NM_0\de+M_0^2)/(1+\de)}$ within the state $\ket{\al,\be}$, and furthermore commutes with $Z$ and hence does not change its value under the transformation \eqref{eq: Ztranslated coherent states}. Hence, the recurrence period within the state $\ket{\al,\be,q}$ can be read off by re-expressing (\ref{TNM0}) through $S^2$, and is given precisely by \eqref{eq: period returns} completing the connection to the classical considerations of section~\ref{sectimeper}.


\section{Discussion}\label{sec:discussion}

We have analyzed the leading order corrections at small coupling to multiparticle energy eigenstates of quantum fields
in AdS$_{d+1}$ with arbitrarily large numbers of particles. The corrections lift the large degeneracies
in the free field energy eigenstates, inducing a fine structure in the spectrum. We have focused on this structure
for the states with a maximal amount of angular momentum for the given energy. Within this sector
we have identified a very large family of analytically tractable states whose energies form simple ladders
\eqref{eq: energy ladder} with explicit wavefunctions that can be read off from \eqref{eq: building blocks ladder states}, \eqref{eq: unnormalised kernel ladder state} and \eqref{eq: general ladder state}. One can furthermore superpose the energy eigenstates in this family to form coherent-like combinations (\ref{eq: definition coherent state})
that recover the special time-periodic solutions present in the corresponding classical dynamics described in section \ref{sectimeper}. In the context of large $c$ holography, these findings imply
explicit infinite families of multiparticle operators in the dual CFT with definite conformal dimensions at order $1/c$, as well as a way to explicitly connect classical weakly nonlinear dynamics in AdS to these operator families.

All of our technical derivations have been phrased for a simple non-gravitating $\phi^4$ scalar in AdS,
but the structure we have presented is robust and should manifest itself in other situations. The technical derivations of sections~\ref{sec:rational} and~\ref{sec:coherent} apply not just to the quantum resonant system (\ref{Hresq}) that controls the energy shifts of the $\phi^4$ probe field, but to any of the resonant systems of the huge class described in \cite{solvable}. Such systems appear naturally for field systems in AdS, provided that there is a breathing
mode in the classical equations of motion (the center-of-mass motion in AdS gives such a breathing mode),
the leading nonlinearities are quartic, and the resonant approximation to the classical dynamics can be consistently truncated to a simple set of modes labelled by one integer \cite{breathing}. The maximally rotating sector of the $\phi^4$ scalar gives a simple implementation of this scenario. The most direct and interesting generalization would be to look for a similar construction involving a scalar field with gravitational or higher-spin interactions in AdS$_3$, adapted to the considerations of \cite{hspin,hspin2}.

Further afield, there are situations where there is a classical consistent truncation resulting in a resonant
system of the class \cite{solvable}, with the associated time periodicities, but it does not translate to an exact truncation of the quantum theory. An example is the spherically symmetric sector of a conformally coupled scalar in AdS$_4$ \cite{CF}. In such situations, one should expect that the pattern we have displayed will not appear in the quantum spectrum exactly, but rather sufficiently high energy levels that start approaching the semiclassical regime will asymptote to our ladder structure. There is furthermore the subject of approximate time periodicities in AdS in the form of Fermi-Pasta-Ulam-like returns \cite{FPU,returns}. An interesting question is what corresponds in the quantum spectra to such approximately periodic weakly nonlinear classical behaviors.

\section*{Acknowledgments}

This work has been supported in part by FWO-Vlaanderen through project G006918N and by Vrije Universiteit Brussel through the Strategic Research Program ``High-Energy Physics.'' MDC is supported by a PhD fellowship from the Research Foundation Flanders (FWO). The work of OE\ is funded by the CUniverse research promotion initiative (CUAASC) at Chulalongkorn University.

\appendix
\section{Asymptotic behavior of the ratio of Jacobi polynomials \eqref{eq: ratio jacobi}}
\label{app: Asymptotic behavior jacobis}
We describe the steps in the computation of the limiting behavior of the ratio of Jacobi polynomials
\begin{align}
 \frac{P_{M-n}^{(1-2(M-n)-\delta (N-1),-1+(N-1)(\delta+1))}(1+2\delta)}{P_M^{(1-2M-\delta N,-1+N(\delta+1))}(1+2\delta)},
 \label{eq: app ratio jacobi}
\end{align}
in the large $M$ and $N$ limit at a fixed ratio $\rho=M/N$ using the tools developed in \cite{jacobi} to approximate a single Jacobi polynomial $P^{(\alpha_m,\beta_m)}_m(x)$ in the large $m$ limit. To this end, we start by summarizing the notation and definitions that were used in \cite{jacobi}. The asymptotics of a Jacobi polynomial can be characterized by the following two parameters:
\begin{equation}
    A \equiv \lim_{m \rightarrow \infty} \frac{\alpha_m}{m}, \qquad B \equiv \lim_{m \rightarrow \infty} \frac{\beta_m}{m}.
\end{equation}
It is straightforward to see that for our purposes $A$ and $B$ are always real and such that $A<-2$, $B>0$ and $A+B>-1$ (e.g.\ for the denominator in \eqref{eq: app ratio jacobi} we find $A=-2-\de \rho^{-1}$ and $B = (\de+1)\rho^{-1}$). In particular, this means that the parameters of interest are outside of the triangle bounded by the lines at $A=0$, $B=0$ and $A+B+2=0$ and the asymptotics of the corresponding Jacobi polynomials has been derived in \cite{jacobi} (to be precise, we will be interested in case C.2 in the notation of \cite{jacobi}).
The large $m$ behavior of the polynomials depends on the position $x$ at which they are evaluated. In our case, we need $x=1+2\de$ and the asymptotic behavior is given by (2.17) in \cite{jacobi}:
\begin{align}
p_m (x) = & \frac{1}{\kappa^m}\,\left(G^m(x) N_{11}(x)
        \left(1 + O\left(\frac{1}{m}\right)\right)\right.  \left. +\,\left(G(x)w^2(x)\right)^{-m}N_{12}(x)
        \left(1 + O\left(\frac{1}{m}\right)\right) \right)\,.
        \label{eq: limits polynomial}
\end{align}
where $p_m (x)$ is the asymptotic behavior for the monic Jacobi polynomial
$$
p_m (x)= \lim_{m \rightarrow \infty} \hat{P}^{(A_m,B_m)}_m(x) = \lim_{m \rightarrow \infty} \frac{2^m m! }{(A_m+B_m+m+1)_m} P^{(A_m,B_m)}_m(x).
$$
The functions (to be evaluated at $x=1+2\de$) and variables appearing in \eqref{eq: limits polynomial} are defined as functions of the parameters $A$ and $B$ and are described in detail in \cite{jacobi}. Here, we simply mention some features that simplify the computation of the ratio \eqref{eq: app ratio jacobi} using \eqref{eq: limits polynomial}. 

We first note that some of the functions appearing in \eqref{eq: limits polynomial} are defined through an integral $\int^x_{\zeta_2} f(t)dt$ for some function $f(t)$ and where $\zeta_2$ is a function of $A$ and $B$. In practice, this lower bound never actually needs to be taken into account because it turns out to cancel in the ratio \eqref{eq: app ratio jacobi}. This statement is actually only true after noticing that the second term in \eqref{eq: limits polynomial} always dominates over the first term in the large $M$, $N$ limit. For this reason, it is sufficient to determine the ratio \eqref{eq: app ratio jacobi} using the second term in \eqref{eq: limits polynomial} with appropriate $A$ and $B$ defined for the polynomial in the numerator and denominator of \eqref{eq: app ratio jacobi}. One then finds in the limit of large $M$ and $N$
 \begin{align}
\frac{P_{M-n}^{(1-2(M-n)-\delta (N-1),-1+(N-1)(\delta+1))}(1+2\delta)}{P_M^{(1-2M-\delta N,-1+N(\delta+1))}(1+2\delta)} =
\frac{(\rho-1) (-1)^{n}  \left( \gamma^{-1}
   \left(\frac{\delta}{\rho}+2\right)+\delta+2\right)^{-n}}{\left( \rho \gamma-1\right) \left((\de +1)^{-1}\gamma^{-1}+1\right)^{\delta}},
\end{align}
with $\rho = M/N$ and $\gamma = \sqrt{(\rho+\de)/(\rho\left(1+\de\right))}$.

\end{document}